\documentclass[sigconf, 10pt]{acmart}
\usepackage{subcaption}
\usepackage{xcolor}
\usepackage{multirow}
\newcommand{\rev}[1]{#1}

\AtBeginDocument{%
  }

\setcopyright{acmlicensed}
\copyrightyear{2018}
\acmYear{2018}
\acmDOI{XXXXXXX.XXXXXXX}
\acmConference[Conference acronym 'XX]{Make sure to enter the correct
  conference title from your rights confirmation email}{
  2018}{Woodstock, NY}
\acmISBN{978-1-4503-XXXX-X/2018/06}

\begin{document}
\title{Breaking Coordinate Overfitting: Geometry-Aware WiFi Sensing for \rev{Cross-Layout} 3D Pose Estimation}

\author{
Songming Jia$^{1\dagger}$,
Yan Lu$^{2\dagger}$,
Bin Liu$^{1*}$,
Xiang Zhang$^{3\dagger*}$,
Peng Zhao$^{4}$,
Xinmeng Tang$^{1}$,
Yelin Wei$^{1}$,
Jinyang Huang$^{4}$,
Huan Yan$^{5}$,
Zhi Liu$^{6}$
}

\affiliation{
$^{1}$ University of Science and Technology of China \country{}
$^{2}$ Shanghai Artificial Intelligence Lab \country{}
$^{3}$ Tianjin University \country{}
$^{4}$ Hefei University of Technology \country{}
$^{5}$ Guizhou Normal University \country{}
$^{6}$ The University of Electro-Communications \country{}.
}

\renewcommand{\shortauthors}{Trovato et al.}

\begin{abstract}

WiFi-based 3D human pose estimation offers a low-cost and privacy-preserving alternative to vision-based systems for smart interaction. However, existing approaches rely on visual 3D poses as supervision and directly regress CSI to a camera-based coordinate system. We find that this practice leads to coordinate overfitting: models memorize deployment-specific WiFi transceiver layouts rather than only learning activity-relevant representations, resulting in severe generalization failures. To address this challenge, we present PerceptAlign, the first geometry-conditioned framework for WiFi-based \rev{cross-layout} pose estimation. PerceptAlign introduces a lightweight coordinate unification procedure that aligns WiFi and vision measurements in a shared 3D space using only two checkerboards and a few photos. Within this unified space, it encodes calibrated transceiver positions into high-dimensional embeddings and fuses them with CSI features, making the model explicitly aware of device geometry as a conditional variable. This design forces the network to disentangle human motion from deployment layouts, enabling robust and, for the first time, layout-invariant WiFi pose estimation. To support systematic evaluation, we construct the largest cross-domain 3D WiFi pose estimation dataset to date, comprising 21 \rev{subjects},  \rev{5 scenes}, 18 actions, and \rev{7 device layouts}. \rev{Experiments show that PerceptAlign reduces in-domain error by 12.3\% and cross-domain error by more than 60\% compared to state-of-the-art baselines.} These results establish geometry-conditioned learning as a viable path toward scalable and practical WiFi sensing.

\end{abstract}




\received{20 February 2007}
\received[revised]{12 March 2009}
\received[accepted]{5 June 2009}

\maketitle

\footnotetext{
$\dagger$~These authors contributed equally to this work.
$^*$~Corresponding authors.
}

\section{Introduction}
\label{sec:intro}

%In recent years, fine-grained 3D human pose estimation is a technical core for many applications, including smart environments, health monitoring, and natural human–computer interaction. Camera-based pipelines achieve high accuracy in controlled settings but raise privacy concerns and fail under occlusion or poor lighting; wearable and radar solutions alleviate some issues but introduce cost, intrusiveness, or deployment constraints. WiFi-based sensing—using commodity Channel State Information (CSI)—offers a promising alternative: it is inexpensive, privacy-preserving, and naturally robust to visual occlusion. Motivated by these advantages, recent work has investigated learning mappings from CSI to 3D pose under visual supervision \cite{PersonInWiFi3D,WiPose,RFPose3D,MMFi}.

%Fine-grained 3D human pose estimation is a technical core for many applications, including smart environments, health monitoring, and natural human–computer interaction. Camera-based pipelines . WiFi sensing has emerged as a compelling alternative. Unlike vision-based systems, WiFi operates without line-of-sight, protects user privacy, and leverages existing infrastructure, making it attractive. Recent studies demonstrate the feasibility of mapping Channel State Information (CSI) to 3D human poses using visual supervision. Yet despite promising in-domain results, these systems collapse when deployed in new rooms, with different devices, or for new users.

\begin{figure*}[ht]
  \centering
  \includegraphics[width=0.85\textwidth]{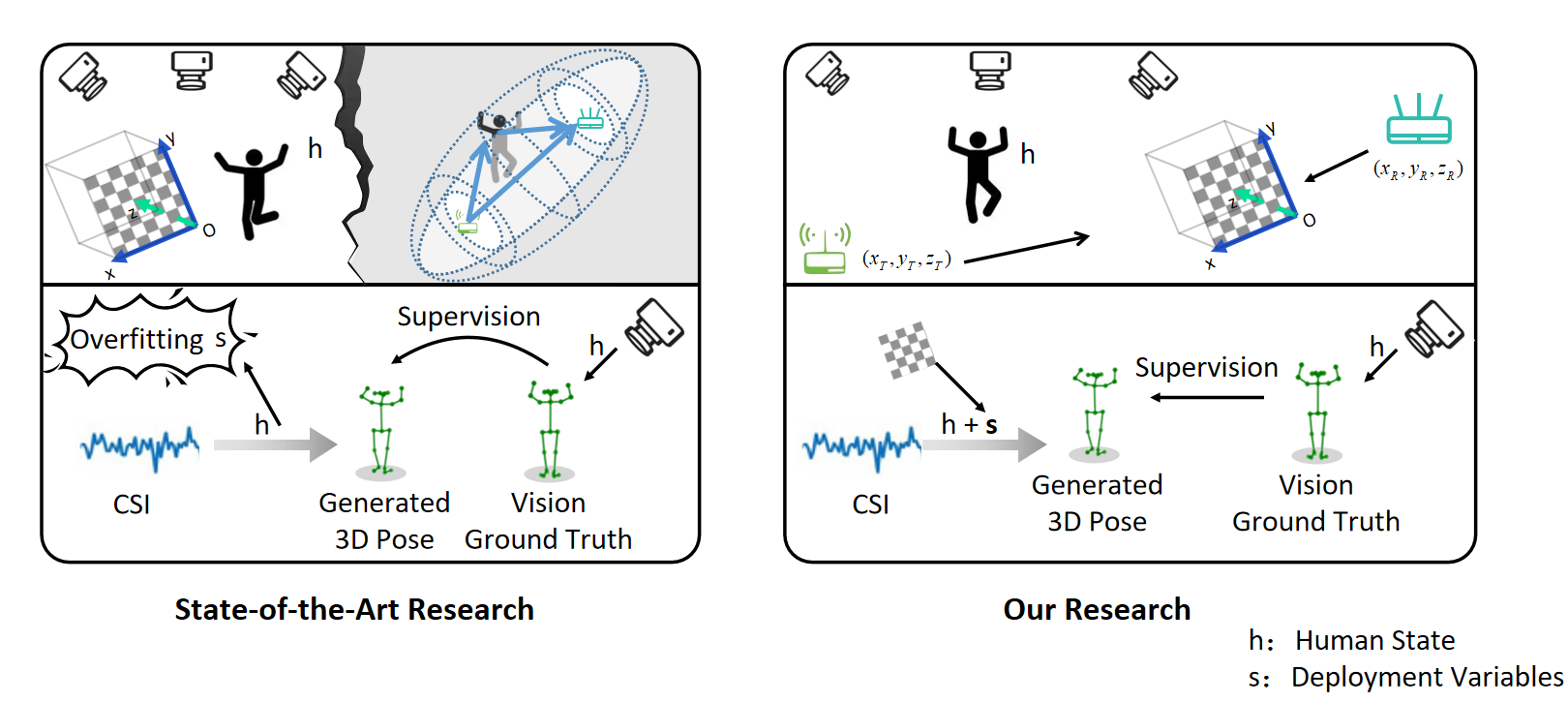}
  %\vspace{-0.1in}
  \caption{\textbf{Left:} Conventional pipelines implicitly memorize the geometric layout of WiFi devices, conflating it with target knowledge and leading to coordinate overfitting.
  \textbf{Right:} PerceptAlign explicitly makes the model aware that WiFi transceiver geometry is a conditional factor rather than knowledge to be memorized.}
  \label{fig:intro_problem}
  %\vspace{-0.1in}
\end{figure*}

In recent years, 3D human pose estimation has served as a core technology for numerous applications~\cite{yeung2025athletepose3d,avogaro2023markerless,lin2023overview}, including health monitoring and natural human–computer interaction. While vision-based methods have long dominated this field~\cite{liu2025tcpformer,neupane2024survey,mehraban2024motionagformer,tripathi20233d}, WiFi sensing has recently emerged as a compelling alternative. Unlike cameras, WiFi sensing operates without line-of-sight, preserves user privacy, and leverages existing infrastructure, making it highly attractive for real-world deployment~\cite{yan2025wi,zhao2024wi,gu2023wife}. %Early studies have demonstrated the feasibility of mapping WiFi Channel State Information (CSI) to 3D skeletons~\cite{yan2024person,jiang2020towards,yang2023mm}. These results established WiFi as a viable sensing modality, sparking a surge of research interest. %Recent studies have demonstrated the feasibility of using WiFi Channel State Information (CSI) for 3D human poses estimation~\cite{yan2024person,jiang2020towards,yang2023mm}.
%The prevailing supervised recipe is straightforward: collect synchronized pairs \((H,\mathbf{y})\), where \(H\) denotes preprocessed CSI and \(\mathbf{y}\) are metric 3D joint coordinates produced by a calibrated visual pipeline; train a network \(\hat{\mathbf{y}}=f_{\theta}(H)\) to minimize Euclidean pose error. This strategy attains competitive in-domain performance when training and test data share the same deployment (room geometry, device placements, hardware). Representative systems following this paradigm include Person-in-WiFi-3D and related RF-pose pipelines \cite{yan2024person,jiang2020towards,yang2023mm,zhou2023metafi++,wang2019can,yang2022metafi,chen2023seeing,yu2023rfpose,zhao2018rf}.
The prevailing WiFi-based pose estimation pipeline is conceptually straightforward. They typically collect synchronized WiFi Channel State Information (CSI) streams together with human poses obtained from calibrated camera systems, and then train a deep network under visual supervision to predict human pose directly from the CSI 
inputs.~\cite{yan2024person,jiang2020towards,yang2023mm,zhou2023metafi++,wang2019can,wang2024xrf55}. Most existing work following this paradigm focuses on improving accuracy and producing smoother pose estimates through tailored loss functions or specialized signal acquisition setups. For example, WiPose~\cite{jiang2020towards} incorporates prior knowledge of human skeleton structure, HPE-Li~\cite{d2024hpe} leverages attention mechanisms, and GoPose~\cite{ren2022gopose} employs a customized antenna array to extract 2D angles of arrival. Other studies push beyond skeleton-level estimation and aim for finer-grained representations, such as reconstructing full 3D human meshes~\cite{wang2024multi} or detailed hand skeletons~\cite{ji2023construct}.

%This progress has established WiFi as a promising sensing modality and sparked a surge of research interest.

These solutions have demonstrated the feasibility of WiFi-based human pose estimation and achieved competitive accuracy when training and testing are performed under the same deployment conditions~\cite{yan2024person,jiang2020towards,yang2023mm,zhou2023metafi++}. However, current methods exhibit striking brittleness when applied to new settings. Although some studies attempt to address environmental (e.g., room) variation through techniques such as multiresolution convolution~\cite{zhang2025wivipose} for richer feature extraction or alignment losses~\cite{zhou2024adapose} for improved supervision, they remain unable to handle the domain shifts frequently encountered in practice, such as changes in transceiver layouts. We argue that the root cause lies in what we term coordinate overfitting. By enforcing a direct regression from CSI to camera generated skeletons, existing pipelines implicitly entangle CSI measurements with deployment-specific WiFi transceiver layouts. As illustrated in Fig.\ref{fig:intro_problem}, visual annotations and WiFi measurements used in conventional methods originate from different coordinate systems. State-of-the-art multi-camera systems generate 3D pose labels in a Cartesian space anchored to a calibration checkerboard, whereas WiFi sensing is inherently tied to the Fresnel zones~\cite{wu2022wifi,zhang2025camlopa,zhang2023wital} defined by transmitter and receiver positions, capturing variations in propagation paths. Training models on CSI using only visual labels as supervision, without accounting for the relationship between WiFi transceiver geometry and the coordinate system of visual ground truth, prevents the model from recognizing that device layout is a conditional variable rather than knowledge to be learned. \rev{As a result, the network is compelled to entangle deployment specific layouts with human pose representations, which prevents effective generalization to unseen layouts and thereby limits its practical applicability~\cite{wang2019can,yang2022metafi,chen2023seeing}.}

To overcome this limitation, we propose PerceptAlign, a geometry conditioned framework that disentangles human motion from deployment-specific artifacts in WiFi-based 3D pose estimation. \rev{Our key idea is straightforward: explicitly exposing device geometry as a varying condition rather than allowing the model to memorize it as hidden background.} Specifically, we first unify heterogeneous WiFi and camera coordinate systems into a shared 3D space through a lightweight coordinate unification procedure. The registered transceiver positions are then encoded as high-dimensional spatial priors and integrated into the learning process, enabling the model to separate human-dependent signals from deployment-dependent factors. This design compels the network to learn features that remain stable under layout changes, ultimately yielding more robust and generalizable WiFi-based 3D pose estimation.
%insight is that visual labels inherently filter out environmental and layout factors, preserving only human motion, whereas WiFi signals encode a composite of human dynamics, transceiver geometry, and multipath reflections. When trained solely with visual supervision, models cannot distinguish which components of CSI correspond to human motion and which stem from deployment conditions. PerceptAlign tackles this issue with a
%PerceptAlign consists of two tightly-coupled components:
%\begin{enumerate}
%  \item \textbf{Board-assisted coordinate unification.} For each device placement we capture the calibration-board pose(s) in camera space and measure device offsets relative to board markers. Composing camera–board and device–board transforms yields transmitter/receiver 3D coordinates \(P=\{\mathbf{p}_{\mathrm{tx}},\mathbf{p}_{\mathrm{rx}_n}\}\) expressed in \(\mathcal{B}\). Recording \(P\) converts the latent transform \(T\) into observed metadata.
 % \item \textbf{Geometry-conditioned learning.} Preprocessed CSI is encoded into image-like feature tensors by a shared CNN encoder. Per-frame CSI features are augmented with compact, multi-scale positional encodings derived from \(P\) (per-receiver spatial tokens) and fed to an attention-based fusion backbone that jointly reasons about spatial and temporal evidence to predict camera-frame 3D joints.
%\end{enumerate}
The workflow of PerceptAlign consists of two main components:

\begin{figure}[ht]
  \centering
  \includegraphics[width=0.45\textwidth]{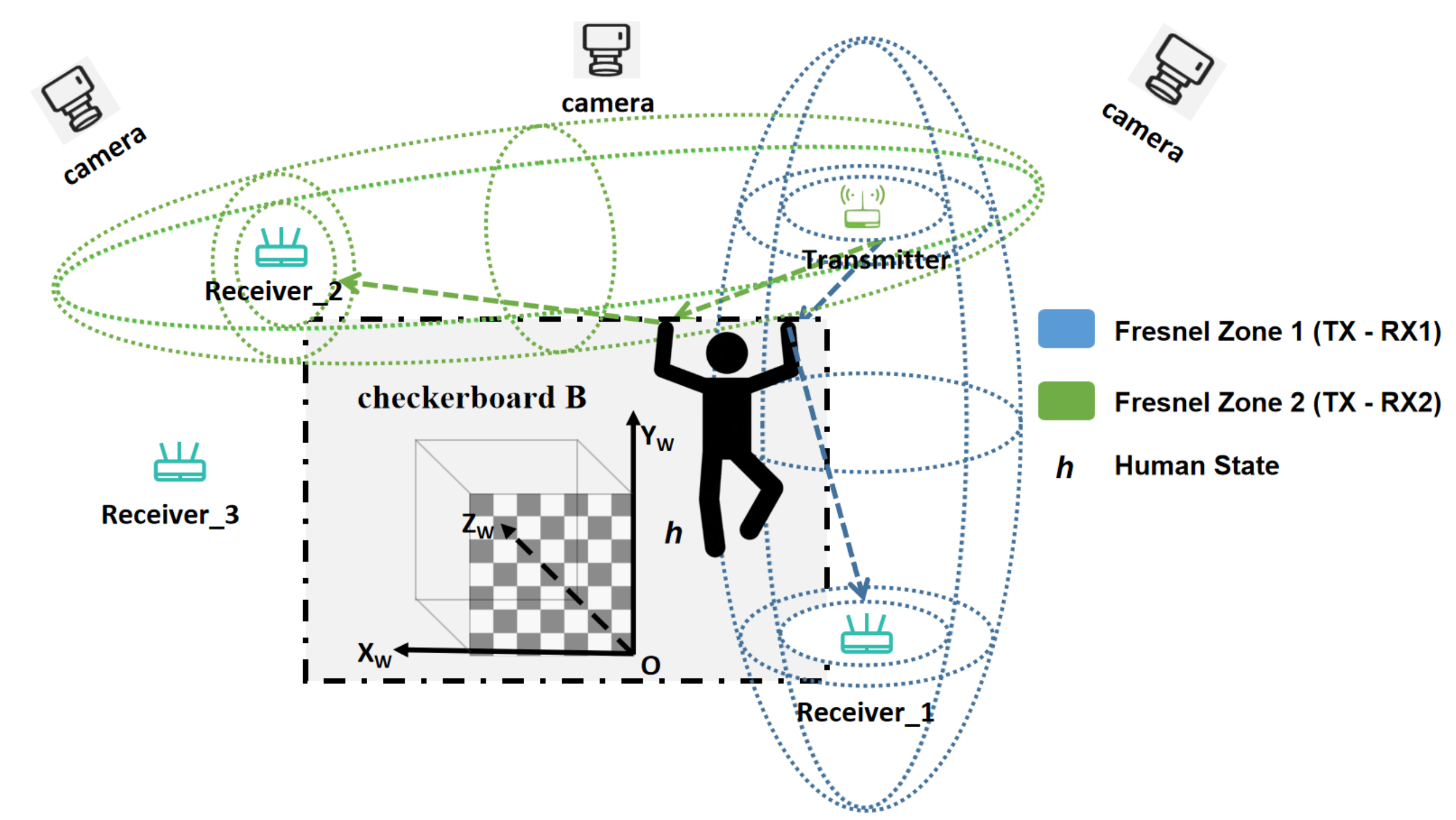} % no spaces in filename
  \caption{Sensing in vision and WiFi. \rev{The Vision system establishes an absolute world coordinate system rooted in the checkerboard $B$. The WiFi system establishes relative sensing coordinate system based on the transceiver locations. The colored ellipsoids represent the Fresnel Zones corresponding to each TX-RX pairs.}}
  \label{fig:Sensing_frames}
  \vspace{-0.15in}
\end{figure}

\textbf{Lightweight Coordinate Unification Procedure.} The goal of this step is to align the coordinate system of WiFi transceivers with that of the multi-camera system, thereby establishing a unified coordinate system. In current multi-camera pose estimation systems, cameras must first capture images of a checkerboard $\mathbb{B}$ placed in the scene and then perform parameter calibration using predefined algorithms in order to obtain accurate 3D human poses. After calibration, all cameras share a world coordinate system defined by the $\mathbb{B}$’s origin and axes, and a transformation matrix is obtained that map the world coordinates into each camera’s imaging coordinate system. After camera calibration, our coordinate unification procedure uses an additional calibration checkerboard $\mathbb{B}_1$ placed between the middle of the WiFi transceiver, with its x-axis aligned along the transceiver line-of-sight (LOS). This checkerboard coordinate system then represents the transceiver coordinate system. Then, we use a camera $\mathbb{C}$ simultaneously observes $\mathbb{B}$ and $\mathbb{B}_1$, allowing us to compute transformation matrices from each checkerboard to the camera. Using the camera as an intermediary, the WiFi transceiver coordinates can thus be mapped into the world coordinate system quickly and conveniently. ($\mathbb{B}_1$->$\mathbb{C}$->$\mathbb{B}$).

\textbf{Geometry-Conditioned Learning.} Once physical-space coordinate unification is established, the WiFi transceiver coordinates are encoded as high-dimensional spatial embeddings. These embeddings are fused with CSI features extracted by a CNN encoder, and the combined representations are processed by an attention-based fusion backbone that jointly reasons about spatial and temporal evidence to predict 3D human poses. By explicitly incorporating transceiver layout as conditional knowledge, the model avoids implicitly memorizing it as background information, thereby mitigating overfitting and substantially improving generalization. And our contributions are as follows:
\begin{itemize}
  \item We reveal coordinate overfitting as the fundamental bottleneck in WiFi-based 3D pose estimation. Existing pipelines directly regress from CSI to camera-generated skeletons, implicitly memorizing deployment-specific layouts, and thus failing to generalize across domains.%A practical cross-modal \emph{coordinate alignment} strategy that registers commodity WiFi transceivers into a calibrated camera frame, converting a major latent deployment degree of freedom into explicit metadata for learning.
  \item We propose PerceptAlign, the first geometry conditioned framework for 3D WiFi pose estimation. It introduces a lightweight coordinate unification procedure that aligns WiFi and vision into a shared space, and a geometry conditioned learning strategy that explicitly encodes transceiver layouts as conditional priors. This design disentangles human motion from device layouts, achieving the first robust WiFi-based pose estimation across transceiver layouts.%The largest synchronized visual–WiFi dataset (multi-scene, multi-configuration, multi-user) with per-sequence device-to-camera calibration released to enable rigorous cross-deployment evaluation.
  \item We construct the largest \rev{cross-domain WiFi-based pose estimation dataset} to date, covering diverse participants, \rev{scenes}, actions, and device setups with detailed geometric calibration. Extensive experiments demonstrate that PerceptAlign reduces in-domain error by \rev{12.3\%} and cross-domain error by more than 60\% over state-of-the-art baselines.%A compact \emph{geometry-conditioned} neural architecture that fuses CSI-derived features with positional embeddings of calibrated device coordinates via attention-based spatial–temporal fusion.
\end{itemize}

%The remainder of the paper motivates and formalizes each component. Sec.~\ref{sec:preliminary} presents empirical evidence and a formal motivation for the misalignment issue; Sec.~\ref{sec:Cross-Modal Coordinate Alignment Strategy} details the calibration and dataset construction; Sec.~\ref{sec:network} describes the geometry-conditioned model and training; Sec.~\ref{sec:Experiments} evaluates accuracy and cross-domain robustness; Sec.~\ref{sec:related} and Sec.~\ref{sec:conclusion} discuss related work and close.

\section{Preliminary}
\label{sec:preliminary}

%In our preliminary study we evaluated a state-of-the-art visual–WiFi pipeline (Person-in-WiFi-3D) on our dataset. While the baseline attains reasonable in-domain 3D pose accuracy under standard train/test splits, its error increases substantially under cross-domain evaluations (e.g., leave-one-scene-out or leave-one-setup-out). These failures point to a structural cause: prior methods overfit WiFi measurements to camera-centric pose labels, causing the learned mapping to memorize dataset-specific device and scene geometry and therefore to generalize poorly. Below we (i) concisely describe the two sensing reference frames (vision and WiFi) and (ii) explain why a direct data-driven mapping from CSI to camera coordinates is ill-posed and leads to coordinate-level overfitting. We conclude by outlining our solution: explicitly unify the frames and provide calibrated device geometry as an input to the learning model.
In this section, we outline how the vision system produces human pose estimates and how the corresponding WiFi sensing system captures motion-related information. We then analyze the overfitting problem in existing approaches and conclude with the motivation for our proposed framework.

\subsection{Vision-based 3D pose estimation}
\label{sec:vision_model}

%Vision-based 3D pose annotation produces metric 3D joint coordinates in a fixed, calibrated world frame. In our setup this frame, denoted \(\mathcal{B}\), is defined by a checkerboard calibration board: a chosen corner is the origin, the board grid directions define the \(x_{\mathcal{B}}\) and \(y_{\mathcal{B}}\) axes, and \(z_{\mathcal{B}}\) is orthogonal to the board plane.

Owing to their unparalleled accuracy and robustness, current SOTA vision-based approaches for 3D human pose estimation typically rely on multi-camera systems. We also use poses generated by such systems serve as high-quality ground truth for our WiFi-based method. As shown in Figure~\ref{fig:Sensing_frames}, this requires the multi-camera system to first calibrate both intrinsic and extrinsic parameters using a checkerboard $\mathbb{B}$. This calibration process involves capturing an image of current scene containing $\mathbb{B}$ and then computing the intrinsic and extrinsic parameters through predefined algorithms.

%A calibrated camera is characterized by an intrinsic parameter matrix:
%\begin{equation}
%K \;=\; \begin{bmatrix} f_x & s & c_x \\ 0 & f_y & c_y \\ 0 & 0 & 1 \end{bmatrix},
%\end{equation}
%together with lens-distortion parameters \(\kappa\) when applicable, and extrinsic parameters \((R,\mathbf{t})\in\mathrm{SO}(3)\times\mathbb{R}^3\) that maps \(\mathcal{B}\)-coordinates into the camera coordinate frame. Under this model a 3D point \(\mathbf{X}_{\mathcal{B}}=[X,Y,Z]^\top\) projects to an ideal image point \(\tilde{\mathbf{x}}=[\tilde u,\tilde v]^\top\) according to
A calibrated camera is characterized by an intrinsic parameter matrix
\begin{equation}
K \;=\; \begin{bmatrix} f_x & s & c_x \\ 0 & f_y & c_y \\ 0 & 0 & 1 \end{bmatrix},
\end{equation}
together with lens-distortion parameters $\kappa$ when applicable. These parameters specify the focal lengths, principal point coordinates, and distortion coefficients of the camera. The calibration also provides extrinsic parameters:
\begin{equation}
(R,\mathbf{t}) \in \mathrm{SO}(3) \times \mathbb{R}^3,
\end{equation}
which define the rotation $R$ and translation $\mathbf{t}$ of each camera relative to a world coordinate system, and SO(3) means special orthogonal group. The world coordinate system is typically established by the checkerboard $\mathbb{B}$: one chosen corner of the board serves as the origin, the grid directions define the $x_{\mathbb{B}}$ and $y_{\mathbb{B}}$ axes, and $z_{\mathbb{B}}$ is set orthogonal to the board plane. With the extrinsic parameters, a camera can project any 3D point $\mathbf{X}_{\mathbb{B}}=[X,Y,Z]^\top$ defined in the world coordinate system to an ideal image point $\tilde{\mathbf{x}}=[\tilde u,\tilde v]^\top$ according to:
\begin{equation}
\begin{aligned}
\mathbf{X}_{c} &= R\,\mathbf{X}_{\mathcal{B}} + \mathbf{t}, \\
\tilde{\mathbf{x}} & = \Pi(\mathbf{X}_{c}) \;=\; 
\begin{bmatrix} f_x\frac{X_c}{Z_c} + c_x \\[4pt] f_y\frac{Y_c}{Z_c} + c_y \end{bmatrix},
\end{aligned}
\end{equation}
and the observed pixel coordinates \(\mathbf{x}\) are related to \(\tilde{\mathbf{x}}\) via distortion model \(\mathbf{x} = \mathcal{D}(\tilde{\mathbf{x}};\kappa)\). Camera calibration \cite{easymocap} estimates \(K,\kappa,R,\mathbf{t}\) from images of the known checkerboard pattern. 

After completing multi-camera calibration, we employ the state-of-the-art EasyMocap framework~\cite{easymocap} to obtain 3D visual human poses. EasyMocap\cite{easymocap} applies multi-view 2D keypoint detection followed by geometric triangulation to obtain 3D skeleton coordinates. Let \(\{\mathbf{x}_i^{(v)}\}_{v=1}^V\) denote the detected 2D location of joint \(i\) in the \(v\)-th calibrated views. EasyMocap recovers the corresponding 3D keypoint \(\mathbf{y}_i\in\mathbb{R}^3\) through triangulation by solving a small linear (or nonlinear reprojection-error) problem:
\begin{equation}
\hat{\mathbf{y}}_i \;=\; \arg\min_{\mathbf{Y}} \sum_{v=1}^V \left\| \mathbf{x}_i^{(v)} - \Pi_v\big(R_v \mathbf{Y} + \mathbf{t}_v; K_v,\kappa_v\big)\right\|^2,
\end{equation}
where \((K_v,R_v,\mathbf{t}_v,\kappa_v)\) are the intrinsics/extrinsics/distortion for view \(v\). The resulting 3D skeleton \(\mathbf{y}=[\mathbf{y}_1^\top,\dots,\mathbf{y}_J^\top]^\top\) is expressed in the world coordinate system.

%Key physical and algorithmic determinants of visual 3D label fidelity include: camera intrinsics \(K\), distortion \(\kappa\), extrinsics \((R_v,\mathbf{t}_v)\) accuracy, imaging resolution, frame rate, and 2D detector noise. Importantly, once calibration is fixed, the visual labels \(\mathbf{y}\) are defined with respect to the \emph{fixed} frame \(\mathcal{B}\); they do not change unless the camera intrinsics/extrinsics or calibration board placement changes.

Clearly, with the aid of the checkerboard, the 3D human poses produced by a multi-camera system are disentangled from environmental and camera layout factors. The system outputs the absolute coordinates of all human skeleton joints in a predefined world coordinate system, thereby providing high-quality ground truth for training WiFi-based pose estimation models.

\subsection{WiFi-based Motion Sensing} 
\label{sec:wifi_model}

%WiFi-based sensing observes electromagnetic propagation between transmitters and receivers and is naturally anchored to the transceiver geometry rather than to a scene-fixed Cartesian origin. We formalize the RF measurement model and the geometric loci that determine spatial sensitivity.
WiFi-based sensing perceives both the environment and human activities by capturing variations in the propagation paths of electromagnetic signals between transmitters and receivers~\cite{li2025enabling,weng2025fm,fan2025sense}. We model the various factors in WiFi-based human pose estimation that influence WiFi signal propagation as follows:
\begin{equation}
\mathcal{G} \;=\; (h,s),
\end{equation}
where \(h\) denotes the dynamic human state and \(s\) denotes the static deployment variables~\cite{han2025rayloc} (TX/RX 3D poses, static scene scatterers, and hardware-specific parameters). Within a synchronized vision–WiFi multimodal sensing setup, the WiFi measurements corresponding to each camera frame are represented as:
\begin{equation}
H_t \in \mathbb{C}^{A\times M\times T},
\end{equation}
where $A$ denotes the number of antenna pairs, $M$ the number of subcarriers, and $T$ the number of sampled time points. For the WiFi sensing system, we define a sensing operator $\mathcal{F}$ that maps the $\mathcal{G}$ to the measured CSI $H_t$:
\begin{equation}
H_t \;=\; \mathcal{F}(\mathcal{G};t)  + \varepsilon, \;=\; \mathcal{F}(h,s;t)  + \varepsilon.
\end{equation}
where $\varepsilon$ models measurement noise and other unmodeled effects. As a result, the measured CSI $H_t$ is a composite signal that inherently couples human motion with deployment-specific device layout geometry and environmental factors. Under a finite-path approximation the $H_t$ is presented as~\cite{liu2023towards,wang2025vr,he2025beam,hu2023muse,hu2025poison}:
\begin{equation}
H(f_m,t) \;=\; \sum_{k=1}^K \alpha_k(h,s,t)\,e^{-j2\pi f_m \tau_k(h,s,t)},
\end{equation}
where \(\tau_k\) and \(\alpha_k\) are path delays and complex amplitudes that depend jointly on \(h\) and \(s\), $f_m$ is the frequency of the $m$-th subcarrier.

According to Fresnel-zone sensing theory~\cite{song2024finersense,zhang2025diffloc}, the impact of object motion on WiFi CSI can be characterized by a Fresnel model in which the transmitter and receiver serve as the focal points of an ellipsoid. As shown in Figure~\ref{fig:Sensing_frames}, the influence of motion on the sensed signal is modeled as variations in the path and distance relative to these two foci. Thus, the coordinate system of WiFi sensing can be geometrically defined by the positions of the transceivers. In WiFi sensing, the absence of intrinsic and extrinsic calibration methods, as found in vision systems, means that there is no unified world coordinate system. As a result, the sensing operator $\mathcal{F}$ is influenced not only by human activity but also by device layout, environmental structures, and measurement noise. Among these factors, changes in transceiver layout significantly alter the propagation paths affected by human motion, creating substantial barriers to generalization. Prior studies~\cite{wu2022wifi,zhang2025wiopen} have also noted that variations in the geometric relationship between device placement and human orientation or position can induce large shifts in the WiFi sensing signal sequences.
%The principal physical determinants of \(\mathcal{F}\) are therefore the deployment and parameters:
%\begin{equation}
%s \owns \{\mathbf{p}_{\text{tx}},\mathbf{p}_{\text{rx}_n}\},\; A,\; \{f_m\}_{m=1}^M,\; B,\; r_{\text{CSI}},\; T,\; \mathcal{S},\; \text{SNR},\ldots
%\end{equation}
%(positions/orientations, antenna/hardware properties, subcarrier set and bandwidth, temporal sampling, static scatterer geometry, and noise characteristics). In particular the device geometry \(\{\mathbf{p}_{\text{tx}},\mathbf{p}_{\text{rx}_n}\}\) anchors Fresnel level sets and nonlinearly reconfigures the mapping \(\mathcal{F}\); modest changes in these coordinates can induce large, non-local changes in the distribution of \(H_t\).

%As illustrated in Fig.~\ref{fig:Sensing_frames}, the camera frame $\mathcal{B}$ and the radio (Fresnel) geometry are distinct. Crucially, these deployment variables \(s\) are \emph{not} the camera calibration parameters that define the visual world frame \(\mathcal{B}\) (camera intrinsics/extrinsics). Consequently, modest changes in device geometry or environment can produce large, non-local shifts in the distribution of \(H\) while the camera-frame pose labels \(\mathbf{y}=\mathcal{T}(h)\) remain unchanged. This asymmetry—strong sensitivity of \(\mathcal{F}\) to \(s\) versus the independence of \(\mathbf{y}\) from \(s\)—is the root cause of dataset-specific memorization by blind CSI→pose regressors.

\subsection{Coordinate Overfitting in WiFi-based 3D Pose Estimation}
\label{sec:misalignment}
Current SOTA WiFi-based 3D pose estimation systems all rely on visual labels for training. The relationship among the actual human pose, the visual annotations, and the observed WiFi CSI can be expressed as follows:
\begin{equation}
\mathbf{y} \;\xleftarrow{\;\mathbb{B}\;} \mathcal{G}
\;\xrightarrow{\;\mathcal{F}\;}\; H,
\end{equation}
where \(\mathbf{y}\) denotes visual 3D annotations (Sec.~\ref{sec:vision_model}), \(\mathcal{G}=(h,s)\) is the complete geometric state with dynamic human component \(h\) and static/deployment component \(s\) (Sec.~\ref{sec:wifi_model}), and \(\mathcal{F}\) is the RF forward operator producing CSI \(H\). From this formulation, we can draw the following conclusion:
\begin{enumerate}
  \item \textbf{Asymmetric dependencies.} Thanks to intrinsic and extrinsic calibration, the visual labels \(\mathbf{y}=\mathcal{T}(h)\) depend (after calibration) only on the dynamic state \(h\), whereas CSI satisfies \(H=\mathcal{F}(h,s)\) and depends jointly on \(h\) and the deployment variables \(s\).
  \item \textbf{Implicit entanglement during learning.} During training, a WiFi-based 3D pose estimation system employs a blind regressor \(f_{\theta}:H\mapsto\mathbf{y}\) that attempts to approximate \(p(\mathbf{y}\mid H)\). This process can be expressed as:
\begin{equation}
  p(\mathbf{y}\mid H) \;=\; \iint p(\mathbf{y}\mid h)\,p(h\mid H,s)\,p(s\mid H)\,\mathrm{d}h\,\mathrm{d}s,
\end{equation}
  the learner must marginalize over the unobserved and varying \(s\). With finite data, the model commonly exploits spurious correlations between \(s\) and \(\mathbf{y}\) in the training set, effectively learning a mapping \(f_{\theta}(H)\approx G_{\theta}(h,s)\) that depends on \(s\).
  \item \textbf{Device-geometry sensitivity.} Although the learning process of a WiFi-based pose estimation model is influenced by all factors in $s$, device geometry is the most sensitive component. According to Fresnel-zone theory, variations in transceiver distance and orientation may cause substantial shifts in the signal propagation patterns induced by the same human motion. Even small changes in transmitter–receiver placements can lead to large, non-local shifts in the distribution of $H$. Since $\mathbf{y}$ remains fixed in $\mathbb{B}$, a regressor that has internalized a deployment-specific alignment will be systematically biased when the transceiver layout changes at test time, resulting in the large in-domain versus cross-domain performance gap observed empirically.
%Device geometry—a principal component of \(s\)—nonlinearly reconfigures Fresnel surfaces and multipath composition; modest changes in TX/RX placements produce large, non-local shifts in the distribution of \(H\). Since \(\mathbf{y}\) is fixed in \(\mathbb{B}\), a regressor that internalized a deployment-specific alignment will be systematically biased when \(s\) changes at test time, producing the large in-domain vs.\ cross-domain performance gap observed empirically.
\end{enumerate}

In summary, while current approaches have achieved promising progress in refining pose estimation through advanced neural architectures and sensing system designs, they remain fundamentally limited by a severe generalization challenge caused by memorizing device geometry. \rev{We refer to this phenomenon as coordinate overfitting.}
%Because the only theoretically robust link between modalities is the shared dependence on \(h\), a practical remedy is to make the dominant components of \(s\) explicit (in particular calibrated device coordinates \(P\)) and condition the model on them. This reduces the marginalization burden on the learner and prevents deployment-specific memorization—an approach we develop in , where the model encodes deployment-specific alignments rather than learning deployment-invariant representations of human motionSec.~\ref{sec:resolution}.

\subsection{Our Motivation}
\label{sec:resolution}
\rev{From the above analysis, the only theoretically robust link between the visual and WiFi modalities in WiFi based pose estimation is \(h\), while the factors in \(s\) act as confounders that are not observable in the visual domain. By explicitly treating the dominant component of \(s\), namely the transceiver geometry, as conditional information analogous to intrinsic or extrinsic parameters, the model can avoid memorizing layout specific patterns that are not transferable. To operationalize this idea, two key steps are required: \textbf{coordinate unification}, which embeds transceiver coordinates into a unified world coordinate system, and \textbf{geometry conditioned learning}, which enforces the use of transceiver geometry as conditional knowledge rather than a learnable knowledge.}

\begin{figure*}[ht]
  \centering
  \includegraphics[width=0.8\textwidth]{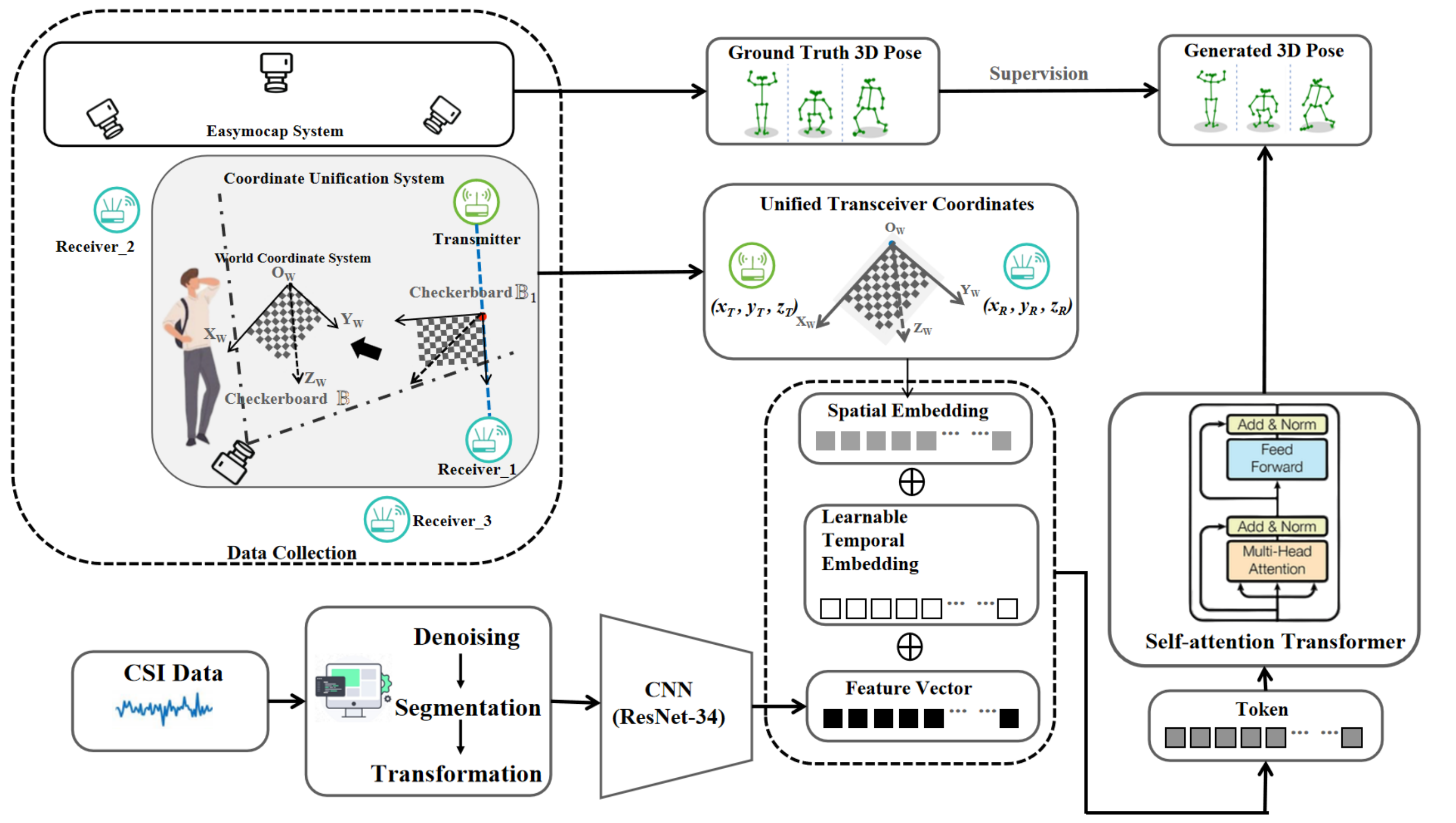}
  %\vspace{-0.1in}
  \caption{\rev{System overview.}}
  \label{fig:system_overview}
  %\vspace{-0.1in}
\end{figure*}

\section{System Overview}
\label{sec:overview}

%PerceptAlign is an end-to-end system that (i) resolves the structural mismatch between WiFi and visual sensing by registering WiFi devices into a camera-centric coordinate frame, and (ii) leverages the unified coordinates to inject explicit geometric priors into a spatially-aware deep model for 3D pose estimation. This section gives a concise technical overview of the two core components: the cross-modal coordinate alignment strategy and the geometry-conditioned neural architecture. Data acquisition and low-level preprocessing steps are described separately in Sec.~\ref{sec:dataset}. Our proposed system is summarized in Fig.~\ref{fig:system_overview}. 
In this section, we present an overview of PerceptAlign, our cross-domain robust WiFi-based 3D pose estimation system. The framework consists of two key components: \textbf{Lightweight Coordinate Unification} and \textbf{Geometry Conditioned Learning}. Specifically, PerceptAlign first performs lightweight coordinate unification in the physical world. This procedure requires only two calibration boards and a few photos, providing a simple and efficient means of aligning coordinate systems. The WiFi sensing system then collects CSI data, which undergoes preprocessing steps including denoising, segmentation, and temporal alignment. A CNN encoder is applied to extract features from the preprocessed data, which are subsequently integrated with high-dimensional embeddings of the WiFi transceiver geometry. The fused representation is finally used to infer the 3D human pose.

\subsection{Lightweight Coordinate Unification.}  
%We employ a practical, board-assisted calibration procedure to compute rigid-body transforms that map transmitter and receiver positions into the camera coordinate frame. The result is a set of calibrated device coordinates expressed in the same Euclidean reference as the visual ground truth. Exposing device geometry explicitly greatly reduces deployment-dependent ambiguity and forms the geometric backbone for subsequent learning. 
The most straightforward approach for coordinate unification would be to manually measure transceiver positions with a ruler and then design a transformation matrix, but this is impractical in real applications due to time and labor requirements. In our method, we introduce an additional checkerboard $\mathbb{B}_1$ to represent the WiFi transceiver coordinate system in the physical world. Using standard camera calibration, we simultaneously establish the transformation from the camera coordinate system to the WiFi checkerboard coordinate system, $T_{C \to B_1}$, as well as the transformation from the camera to the world coordinate system, $T_{C \to B}$. \rev{Using the camera as an intermediary, the mapping from the WiFi coordinate system to the world coordinate system can be efficiently established. Consequently, the WiFi transceiver coordinates can be expressed in the world coordinate system.} % as
%\begin{equation}
%T_{B_1\to B} \;=\; T_{C\to B}^{-1}\,T_{C\to B_1},
%\end{equation}

%Full calibration algorithms, error analysis, and storage/metadata conventions are deferred to Sec.~\ref{sec:Cross-Modal Coordinate Alignment Strategy}.

\subsection{Geometry-Conditioned Learning.}  
%On top of the unified coordinates, the model augments CSI-derived features with compact spatial embeddings of receiver/transmitter locations and applies a lightweight fusion backbone to jointly reason about spatial and temporal evidence. In brief, per-frame CSI features are combined with positional encodings of calibrated device coordinates and processed by a Transformer-based fusion module to produce per-frame 3D joint estimates. Training supervision, loss choices, and architectural hyperparameters are given in Sec.~\ref{sec:network}.
%On top of the unified coordinates, the model augments CSI-derived features with compact spatial embeddings of receiver/transmitter locations (device geometry \(P\)) and applies a lightweight fusion backbone to jointly reason about spatial and temporal evidence.
For geometry-conditioned learning, we replace the blind regressor by a conditioned estimator:
  \begin{equation}
  \hat{\mathbf{y}} \;=\; g_{\theta}\big(H,P\big),
  \end{equation}
where \(P\) reduces the uncertainty about \(s\). Conditioning on \(P\) changes the posterior from \(p(\mathbf{y}\mid H)\) to \(p(\mathbf{y}\mid H,P)\), which is typically much sharper and more stable when \(P\) captures the dominant geometric conditions.  Operationally, this reduces the need for the network to memorize deployment-specific transforms and yields markedly improved cross-deployment generalization. Specifically, we augments CSI features with compact spatial embeddings of receiver/transmitter locations and applies a lightweight fusion backbone to jointly reason about spatial and temporal evidence.%In practice we encode \(P\) via multi-scale positional embeddings and inject these spatial tokens into a Transformer fusion backbone along with CSI features; the network thus learns to interpret CSI in the explicit context of the calibrated device layout.

\section{Method}
\label{sec:metho}

%In this section, we present the methodology of our proposed system, PerceptAlign, an end-to-end framework for robust and generalizable 3D human pose estimation using commercial WiFi devices. The system is meticulously designed to overcome the fundamental challenges discussed in the introduction by integrating novel solutions at three key stages: data collection, data preprocessing, and a sophisticated deep learning network for pose estimation. In the following subsections, we detail each of these components and highlight the unique contributions that enable our system to achieve superior performance and generalization.
In this section, we provide a detailed description of the implementation of the PerceptAlign.

\subsection{Lightweight Coordinate Unification}
\label{sec:Cross-Modal Coordinate Alignment Strategy}
%The core of our approach lies in addressing the fundamental misalignment between visual and WiFi sensing modalities. Existing multi-modal methods suffer from a "perspective overfitting" problem because they fail to account for the distinct coordinate systems used by each modality. This subsection details our innovative strategy to unify these two perspectives, a process that is essential for bridging this gap and enabling a spatially-aware deep learning model.

%The central practical step of our alignment pipeline is to obtain metric 3D coordinates for WiFi transmitters and receivers in the global calibration-board frame \(B\). 
%We achieve this by (i) observing an auxiliary board frame \(B_1\) that is collocated with the device pair, (ii) expressing device offsets in \(B_1\) using measured inter-device separation, and (iii) composing the camera\(\to\)board transforms to map the device coordinates into the global frame \(B\).

%\begin{figure*}[t]
%    \centering
%    \includegraphics[width=\linewidth]{figure/Coordinate_unification_flowchart.pdf}
%    \caption{\rev{Schematic of the lightweight coordinate unification process. }}
%    \label{fig:unification_flowchart}
%\end{figure*}

The goal of this component is to conveniently transform the 3D spatial coordinates of WiFi transceivers into the world coordinate system $B$ defined by the checkerboard $\mathbb{B}$. We achieve this by:(i) introducing an additional checkerboard $\mathbb{B}_1$ to align the sensing coordinate system $B_1$ of the WiFi transceiver pair, (ii) expressing transceiver device offsets in $B_1$ using measured inter-device distance, and (iii) capturing both $\mathbb{B}$ and $\mathbb{B}_1$ with a camera to obtain the corresponding transformation matrices, then using the camera as an intermediary to map WiFi transceiver coordinates from $B_1$ to $B$. \rev{An example of coordinate unification for the TX–RX1 pair is illustrated in the upper left of Figure~\ref{fig:system_overview}. }

%For completeness we repeat the notation: for a point expressed in frame \(X\) its Euclidean coordinate is \(\mathbf{p}_X\in\mathbb{R}^3\) and its homogeneous form \(\tilde{\mathbf{p}}_X=[\mathbf{p}_X^\top\; 1]^\top\). A rigid transform from \(X\) to \(Y\) isWe place the checkerboard \(\mathbb{B}_1\) such that the orientation of its coordinate frame is aligned with the TX–RX1 direction and position it between the transmitter and receiver. In this way, \(\mathbb{B}_1\) encodes the spatial configuration of the transceiver pair. By capturing an image that contains both \(\mathbb{B}\) and \(\mathbb{B}_1\), we estimate the geometric mapping between the two coordinate systems, which in turn enables the transceiver coordinates to be transformed into the world coordinate system.
We begin with the following notation: for a point expressed in coordinate system $X$, its Euclidean coordinate is denoted as $\mathbf{p}_X \in \mathbb{R}^3$, and its homogeneous form is $\tilde{\mathbf{p}}_X = [\mathbf{p}_X^\top \; 1]^\top$. The homogeneous form vectorizes the 3D coordinate, allowing it to be manipulated in matrix form. The transformation of coordinates between reference frames using extrinsic parameters, i.e., a rigid transform, can be expressed as:
\begin{equation}
\mathbf{T}_{X\to Y} \;=\;
\begin{bmatrix}
\mathbf{R}_{X\to Y} & \mathbf{t}_{X\to Y} \\
\mathbf{0}^\top & 1
\end{bmatrix}\in\mathrm{SE}(3),
\end{equation}
with \(\tilde{\mathbf{p}}_Y=\mathbf{T}_{X\to Y}\,\tilde{\mathbf{p}}_X\).

%We place the origin of \(B_1\) at the midpoint of the line joining a transmitter \(T\) and a receiver \(R\) (or at a convenient marker near that midpoint) and align the \(x\)-axis of \(B_1\) with the \(T\!-\!R\) direction. Let the physical (Euclidean) distance between \(T\) and \(R\) be denoted by \(S\) (in metres). Writing \(L=S/2\) for the half-separation, the device coordinates in \(B_1\) are
In the physical setup, after the multi-camera system has been calibrated using $\mathbb{B}$, we deploy an additional checkerboard $\mathbb{B}_1$ between the WiFi transceivers. We place the origin of $\mathbb{B}_1$ at the midpoint of the line joining a transmitter $T$ and a receiver $R$ (or at a convenient marker near that midpoint), and align the $x$-axis of $\mathbb{B}_1$ with the $T\!-\!R$ direction. In this way, the checkerboard coordinate system coincides with that of the WiFi sensing setup. Let the physical Euclidean distance between $T$ and $R$ be denoted as $S$ (in meters). Writing $L = S/2$ for the half-distance, the device coordinates (offsets) in $B_1$ can then be expressed as:
\begin{equation}\label{eq:B1_coords_meters}
\mathbf{p}_{B_1,T} = \begin{bmatrix}-L\\[4pt]0\\[4pt]0\end{bmatrix}, \qquad
\mathbf{p}_{B_1,R} = \begin{bmatrix}L\\[4pt]0\\[4pt]0\end{bmatrix}.
\end{equation}

In practice \(S\) may be obtained either by direct physical measurement (tape measure) or by image-based measurement using the auxiliary camera that images the checkerboard. If the distance is measured in board grid units (for example \(g\) checkerboard squares) and each square has side length \(d\) (in metres), then \(S=g\cdot d\) and the half-distance is \(L=\tfrac{g d}{2}\). Equivalently, if the pixel distance between device markers is \(p\) and the pixel-to-metre scale on the checkerboard is \(\rho\) (meters per pixel, obtained from board calibration), then \(S=p\cdot\rho\) and \(L=\tfrac{p\rho}{2}\). These conversion formulae give a single, implementation-ready rule:
\begin{equation}
L \;=\; \frac{1}{2}\cdot\bigg(\text{measured distance}\bigg)
\;=\; \frac{1}{2}\cdot
\begin{cases}
S & (\text{direct})\\[4pt]
g\,d & (\text{grid units})\\[4pt]
p\,\rho & (\text{pixels})
\end{cases}.
\end{equation}

\begin{figure*}[t]
  \centering
  \includegraphics[width=1\textwidth]{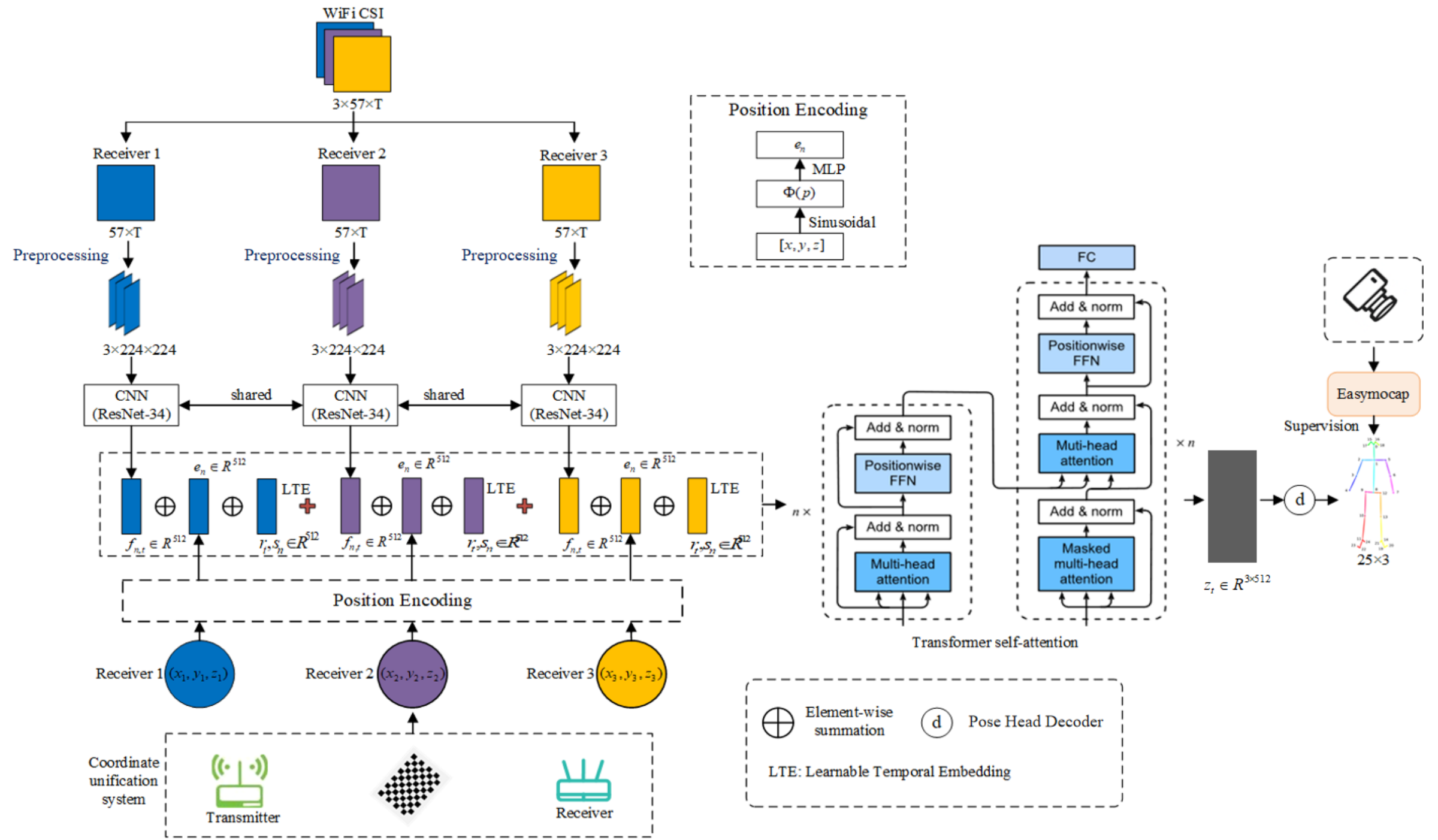}
  %\vspace{-0.1in}
  \caption{\rev{Geometry-conditioned learning network architecture.}}
  \label{fig:network_arch}
  \vspace{-0.1in}
\end{figure*}

Let $\mathbb{C}$ denote the auxiliary camera coordinate system that observes both $\mathbb{B}$ and $\mathbb{B}_1$. From checkerboard calibration we obtain camera-to-checkerboard transforms \(\mathbf{T}_{C\to B}\) and \(\mathbf{T}_{C\to B_1}\). The rigid transform that maps coordinates expressed in \(B_1\) into the world coordinate
system \(B\) is
\begin{equation}\label{eq:T_B1_to_B_repeat}
\mathbf{T}_{B_1\to B} \;=\; \mathbf{T}_{C\to B}^{-1}\,\mathbf{T}_{C\to B_1}.
\end{equation}
Therefore the device homogeneous coordinates in the world coordinate
system \(B\) are obtained by:
\begin{equation}
\tilde{\mathbf{p}}_{B,\alpha} \;=\; \mathbf{T}_{B_1\to B}\,\tilde{\mathbf{p}}_{B_1,\alpha},
\qquad \alpha\in\{T,R\},
\end{equation}
or in non-homogeneous form
\begin{equation}
\mathbf{p}_{B,\alpha} \;=\; \mathbf{R}_{B_1\to B}\,\mathbf{p}_{B_1,\alpha} + \mathbf{t}_{B_1\to B}.
\end{equation}

The procedure above is applied for every transceiver pair. If multiple receivers are observed simultaneously in a single auxiliary view, their positions can be read off in the same \(B_1\) coordinate system and converted en masse via \eqref{eq:T_B1_to_B_repeat}. The result is a set of calibrated device coordinates expressed in the unified world coordinate system:
\begin{equation}
P \;=\; \{\mathbf{p}_{B,\text{tx}},\,\mathbf{p}_{B,\text{rx}_1},\,\mathbf{p}_{B,\text{rx}_2},\dots\}.
\end{equation}

We summarize the implementation summary as follows.
\begin{enumerate}
  \item Place the auxiliary board $\mathbb{B}_1$ so its origin is at the midpoint of the device pair and align its \(x\)-axis with the device line.
  \item Acquire images of $\mathbb{B}_1$ with the auxiliary camera and detect board corners to obtain \(\mathbf{T}_{C\to B_1}\); likewise obtain \(\mathbf{T}_{C\to B}\).
  \item Measure the inter-device distance in board-grid units \(g\) (so \(S=g d\)), or in pixels \(p\) (so \(S=p\rho\)); compute \(L=S/2\).
  \item Form \(\mathbf{p}_{B_1,\{T,R\}}\) via \eqref{eq:B1_coords_meters} and transform to the world coordinate system using \eqref{eq:T_B1_to_B_repeat}.
  \item Repeat for all devices to obtain the full set \(P\) expressed in \(B\).
\end{enumerate}

%By explicitly measuring \(P\) in the world coordinate system \(B\), we convert the previously latent deployment transform into observed metadata, allowing downstream learning modules to condition on the actual device geometry (Sec.~\ref{sec:resolution}).The calibration board-assisted coordinate alignment strategy is designed to bridge the gap between these two disparate coordinate systems. This strategy ensures that the WiFi perception system is spatially "visible" to the visual domain, thereby enabling a unified understanding of space across modalities. 
By explicitly measuring $P$ in the world coordinate system $B$, we make the previously invisible device geometry perceptible to the vision-based annotation system. This enables transceiver coordinates to be introduced as conditional knowledge within a unified reference frame, thereby guiding the model to learn deployment-invariant features. Our proposed lightweight unification strategy requires only the placement of two checkerboards and the capture of a few photos, allowing cross-modal coordinate alignment to be completed efficiently within a short time.
\qedhere

\subsection{Geometry-conditioned learning}
\label{sec:network}

We implement the geometry-conditioned learning illustrated in Fig.~\ref{fig:network_arch}. Each receiver's CSI tensor is encoded by a shared CNN; calibrated coordinates $P$ are encoded via multi-band sinusoidal features and an MLP to form spatial tokens that are concatenated with CNN features. A Transformer-based spatio-temporal encoder fuses these tokens across receivers and time, and a lightweight decoder outputs camera-frame 3D skeleton joints.

\textbf{Preprocessing.}
For each receiver \(R_n\) we compute the complex \emph{CSI ratio} between antenna 1 and antenna 2 for denoising~\cite{zeng2019farsense}:
\begin{equation}
\tilde{c}_{n}(t,f) \;=\; \frac{c_{n,1}(t,f)}{c_{n,2}(t,f)},
\end{equation}
where \(c_{n,a}(t,f)\) denotes the CSI at receiver \(n\), antenna \(a\), time index \(t\) and subcarrier \(f\). The ratio eliminates amplitude and phase noise introduced by hardware. After denoising, the CSI stream is segmented and temporally synchronized with each camera frame to serve as the network input. Specifically, we first divide the CSI sequence into several groups, where each group sequentially contains $G = \lfloor N_c / N_f \rfloor$ CSI samples, with $N_c$ denoting the total number of CSI samples and $N_f$ the number of video frames. Thus, each frame is paired with a fixed-length CSI group. From each group, we extract magnitude, phase, and Doppler Frequency Shift (DFS)~\cite{niu2022rethinking} along the temporal axis. These features are concatenated following the procedure in WiGRUNT~\cite{gu2022wigrunt,zhang2025wiopen}, resized to $1 \times 224 \times 224$, and then replicated across three channels to form a $3 \times 224 \times 224$ tensor. This process is repeated for all WiFi transceivers to construct the CSI inputs corresponding to the visual annotations.

%We aggregate the raw CSI samples into synchronized frame-level inputs by grouping every \(T=\lfloor S/L\rfloor\) consecutive CSI samples (where \(S\) is the total number of CSI samples and \(L\) is the number of video frames) and assigning each group to its corresponding video frame, so that each video frame is paired with a fixed-size CSI segment. From each group we extract magnitude and phase and compute differential features (DFS) along the temporal axis to highlight short-term changes. Each resulting single-channel DFS map is resized to \(1\times 224 \times 224\) and then replicated three times to form a \(3\times 224 \times 224\) tensor compatible with ImageNet backbones. Repeating this for the three receivers yields three image-like inputs per frame.

\textbf{Shared CNN encoder.}
%The three \(3\times224\times224\) inputs are passed through a shared-weight CNN encoder \(E_\theta\), implemented as a pretrained ResNet-34 truncated before the final classification layer. Applied frame-wise, the encoder followed by a pooling head produces per-receiver, per-frame feature vectors%\rev{To extract feature representations from the CSI streams, we employ a weight-sharing Siamese network architecture. Specifically, a shared backbone encoder $E_{\theta}$, implemented using the ResNet-34 architecture, is applied independently and in parallel to the input tensor $X_{n,t}$ of each receiver $n$ at frame index $t$. It is important to note that the encoder parameters $\theta$ are tied and identical across all receiver branches. In terms of implementation specifics, we extract the feature maps directly from the final convolutional stage of the ResNet-34, strictly before the global average pooling layer. This design choice preserves the spatial dimensions of the feature maps, which is crucial for the subsequent fusion with geometric embeddings.}
\rev{CSI input from all receivers is processed through a shared CNN encoder $E_\theta$ to extract features. Specifically, the CSI streams collected at the three receivers are fed into the same encoder $E_\theta$, which is fine-tuned across all inputs.} In this work, $E_\theta$ is implemented as a pretrained ResNet-34 truncated before the final classification layer. It can be expressed as:
\begin{equation}
\mathbf{f}_{n,t} \;=\; \mathrm{Pool}\big(E_\theta(\mathbf{X}_{n,t})\big) \in \mathbb{R}^D,
\end{equation}
\rev{where $\mathbf{f}_{n,t}$ is the feature, which is taken from the output of the final convolutional layer of the CNN encoder.} \(\mathbf{X}_{n,t}\) denotes the input tensor for receiver \(n\) at frame \(t\), \(n\in\{1,\dots,N_r\}\), and \(t\in\{1,\dots,T\}\).

\textbf{Position encoding.}
%All device coordinates and ground-truth poses are expressed in a transmitter-centric frame to remove transmitter-dependent offsets. Let \(\mathbf{p}_n=(x_n,y_n,z_n)\in\mathbb{R}^3\) be the 3D position of receiver \(R_n\) in this frame, and let \(\mathbf{y}_{t}\in\mathbb{R}^{J\times 3}\) denote the ground-truth joint positions at frame \(t\).
%Each receiver coordinate \(\mathbf{p}_n\) is lifted to a high-dimensional spatial embedding via a multi-frequency mapping \(\Phi:\mathbb{R}^3\to\mathbb{R}^{D_p}\):
We represent the geometry of each antenna pair using the coordinate offset of the receiver relative to the transmitter in the world coordinate system. Let $\mathbf{p}_n = (x_n, y_n, z_n) \in \mathbb{R}^3$ denote the 3D geometric relation of receiver $R_n$. Each receiver coordinate $\mathbf{p}_n$ is then lifted into a high-dimensional spatial embedding through a multi-frequency mapping $\Phi: \mathbb{R}^3 \to \mathbb{R}^{D_p}$:
\begin{equation}
\Phi(\mathbf{p}) \;=\; \Big[ \sin(2^{k}\pi \mathbf{p}),\ \cos(2^{k}\pi \mathbf{p}) \Big]_{k=0}^{K-1},
\end{equation}
\rev{Here, $K$ denotes the number of frequency bands utilized for spectral expansion, and $D_p$ represents the dimensionality of the resulting coordinate embedding vector ($K=10$ in this paper, and  $D_p = 6K$ for 3D coordinates). This vector is subsequently projected by a small MLP $g_{\psi}$ to obtain the final spatial embedding}
%and projected by a small MLP \(g_\psi\) to obtain the spatial embedding
\begin{equation}
\mathbf{e}_n \;=\; g_\psi\big(\Phi(\mathbf{p}_n)\big) \in \mathbb{R}^D.
\end{equation}
This step is designed to enrich the geometric representation by expanding the $\mathbf{p}_n$ across multiple frequencies, enabling the model to capture more complex patterns. It also prevents the model from relying solely on shallow features such as simple relative displacements, and further encodes the geometry into an embedding space with a dimensionality comparable to CSI features, which facilitates effective feature fusion.

\textbf{Temporal encoding.}
To represent temporal order we associate each frame \(t\) with a learnable temporal embedding (\rev{LTE}) \(\mathbf{r}_t\in\mathbb{R}^D\) (shared across receivers). \rev{Specifically, we initialize a parameter matrix $\mathbf{R} \in \mathbb{R}^{T_{seq} \times D}$, where $T_{seq}$ denotes the temporal window size and $D$ is the embedding dimension. For each time step $t$, the corresponding vector $\mathbf{r}_t$ provides a unique, dense positional signature. During training, these vectors are optimized via backpropagation to encode relative temporal dependencies, thereby enabling the self-attention mechanism to distinguish and order sequential motion cues. A small, learnable receiver-specific bias $\mathbf{s}_n \in \mathbb{R}^D$ is also added to capture static hardware idiosyncrasies.}%A small, receiver-specific bias \(\mathbf{s}_n\in\mathbb{R}^D\) is also maintained to capture per-receiver idiosyncrasies.

\begin{figure*}[ht]
  \centering
  \begin{subfigure}[b]{0.3\linewidth}
    \centering
    \includegraphics[width=\linewidth]{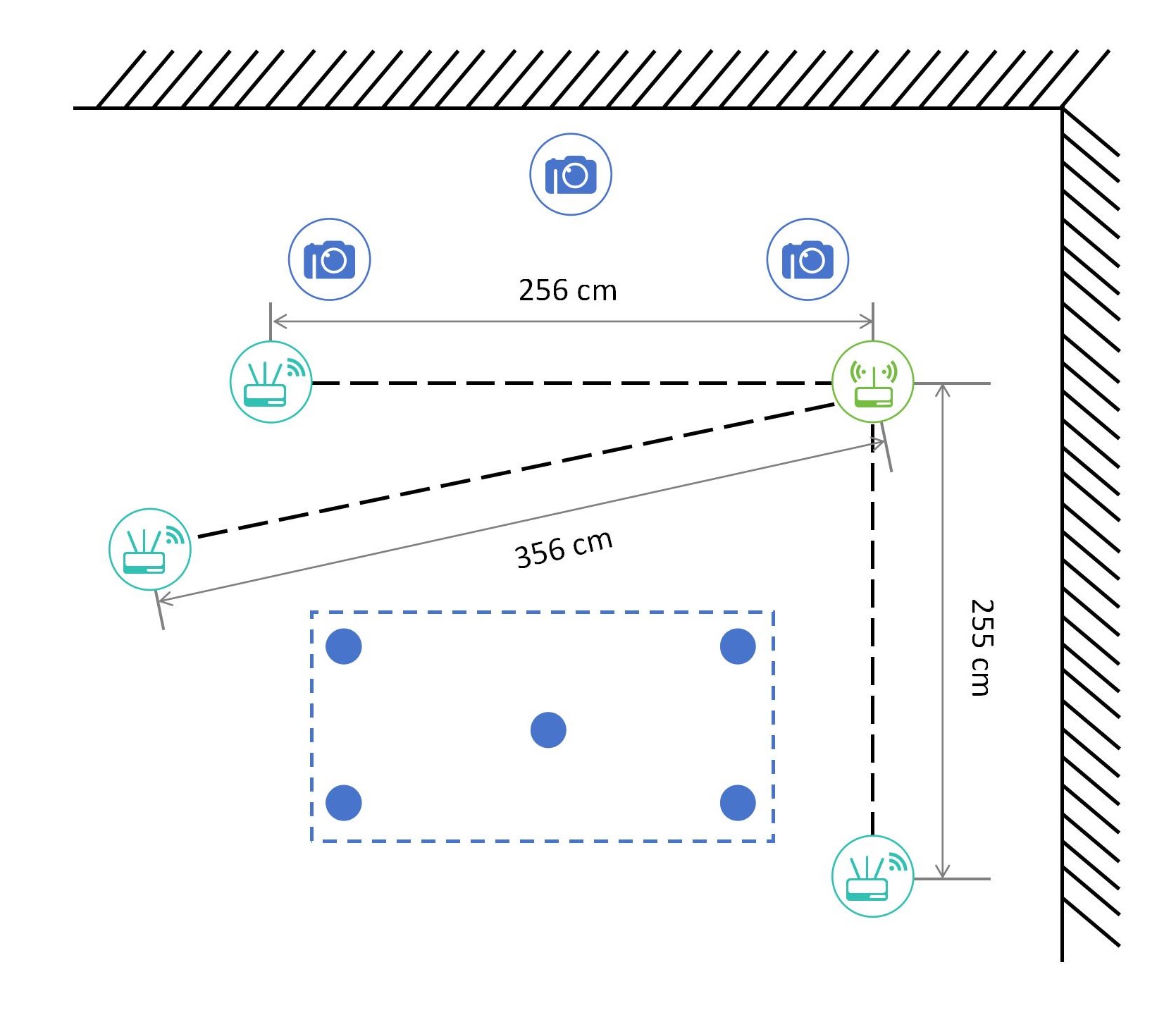}
    \caption{Scene 1: Empty room)}
  \end{subfigure}
  \hfill
  \begin{subfigure}[b]{0.3\linewidth}
    \centering
    \includegraphics[width=\linewidth]{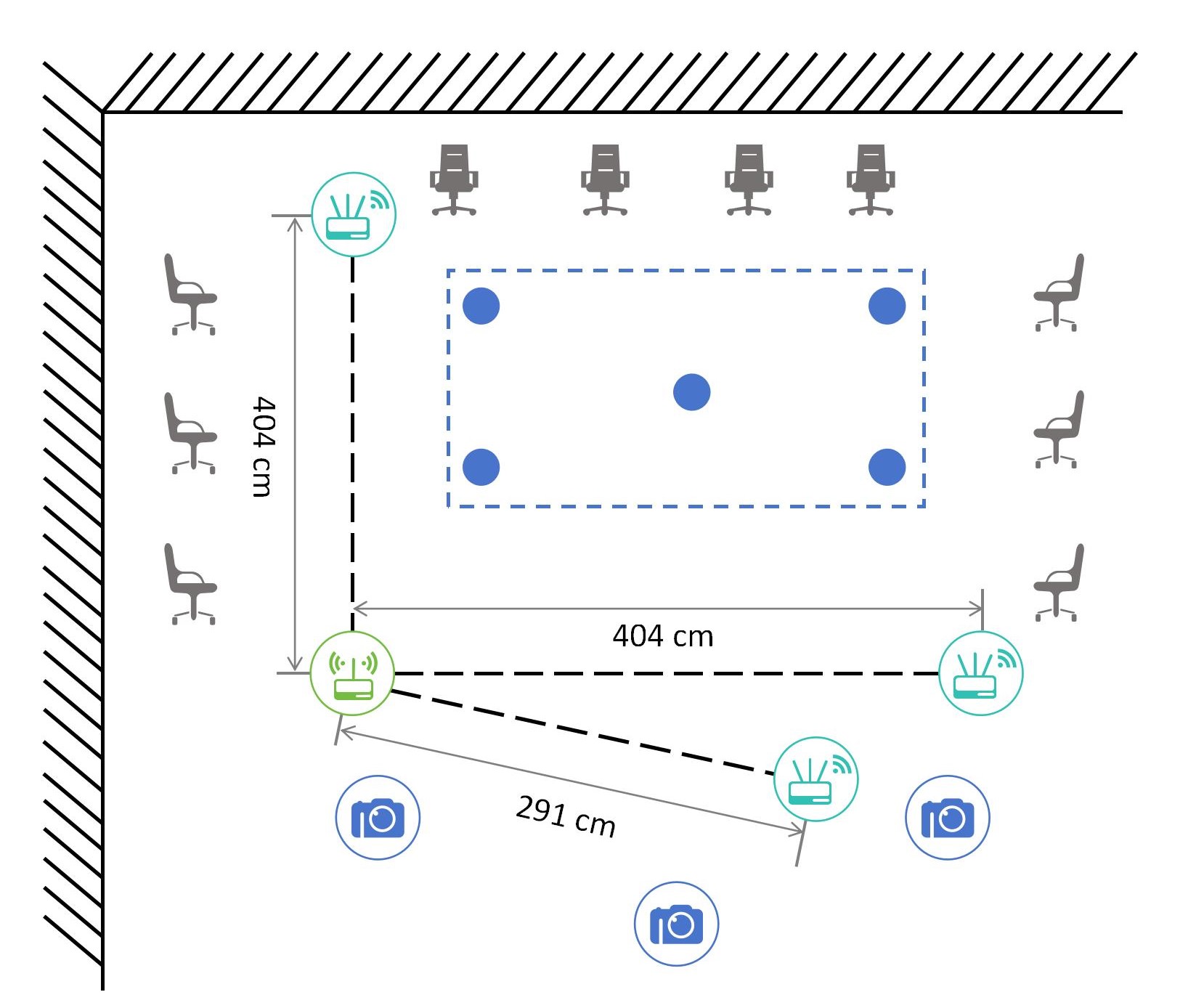}
    \caption{Scene 2: Meeting room)}
  \end{subfigure}
  \hfill
  \begin{subfigure}[b]{0.29\linewidth}
    \centering
    \includegraphics[width=\linewidth]{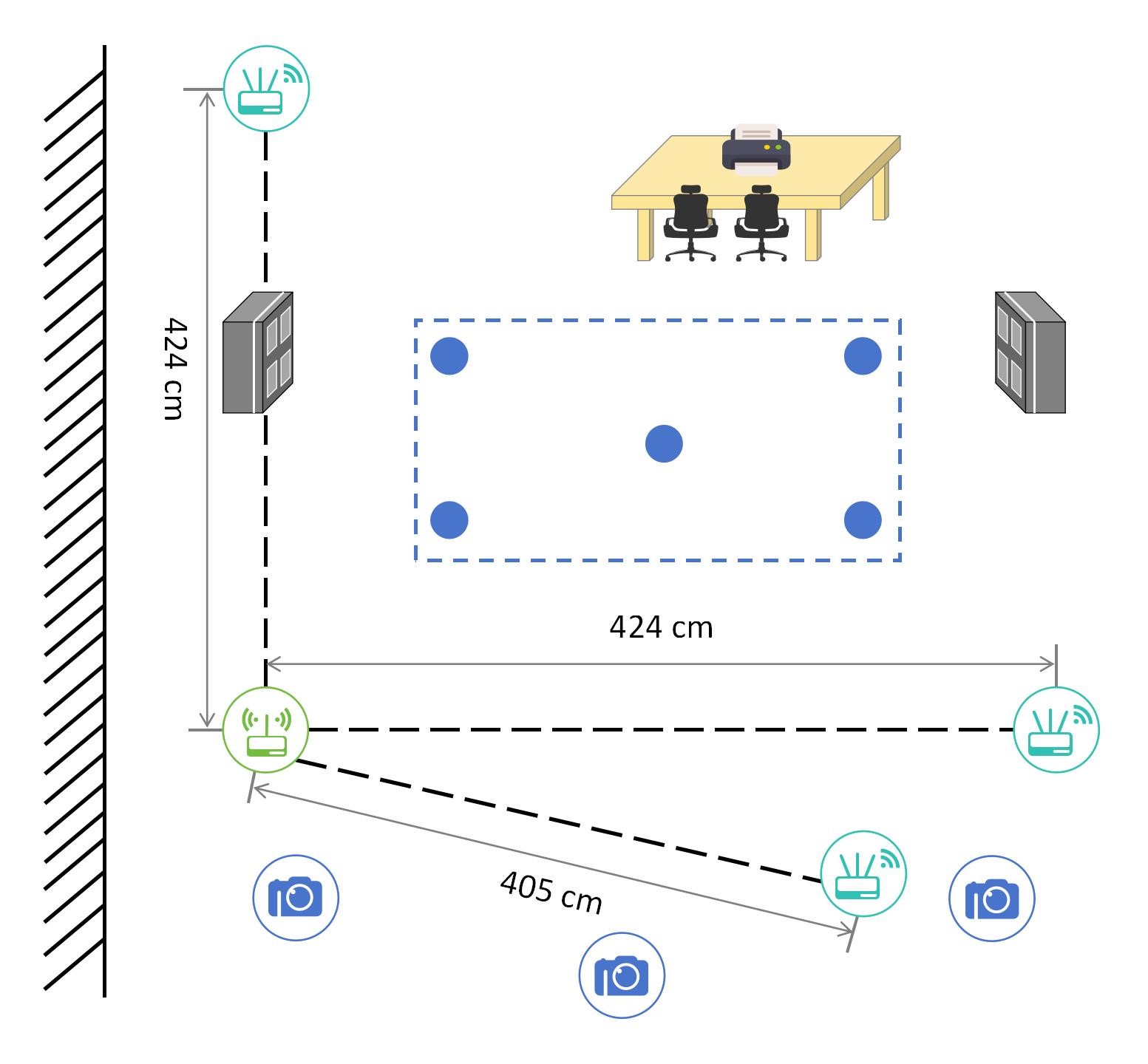}
    \caption{Scene 3: Office}
  \end{subfigure}
%\vspace{-0.1in}
  \caption{\rev{Scenes} and device placements used for data collection. Scene\_3 contains multiple layouts (A/B/C) as documented in the dataset README file.}
  \label{fig:scenes}
  %\vspace{-0.15in}
\end{figure*}

\textbf{\rev{Token construction.}}
\rev{To integrate the learned features and the spatial-temporal embedding, we employ a projection-and-summation fusion strategy. Specifically, \(\mathbf{f}_{n,t}\) and the spatial embedding \(\mathbf{e}_n\) are first projected into the same dimension. They are then combined with \(\mathbf{r}_t\) and \(\mathbf{s}_n\) via element-wise summation to form the tokens, as formulated below:}
\begin{equation}
\mathbf{u}_{n,t}^{(0)} \;=\; \mathrm{LayerNorm}\big(\mathbf{W}_f\mathbf{f}_{n,t} + \mathbf{W}_e\mathbf{e}_n + \mathbf{r}_t + \mathbf{s}_n\big)\in\mathbb{R}^D,
\end{equation}
where \(\mathbf{W}_f,\mathbf{W}_e\) are projection functions. \rev{The final input to the Transformer encoder is formed by concatenating the tokens extracted from each receiver, ordered according to the receiver index and the temporal sequence:}
\begin{equation}
\mathcal{U}^{(0)} \;=\; \big\{ \mathbf{u}_{n,t}^{(0)} \big\}_{n=1,t=1}^{N_r,T},
\end{equation}
of length \(N_r * T\). In the final token, the spatial embedding provides geometric priors of the transceivers, while the temporal embedding enables the attention layers to condition temporal evidence on explicit spatial priors at each timestep. This design encourages the model to learn time-varying observation-to-geometry mappings, leading to improved temporal consistency and higher pose estimation accuracy.%Although the spatial embedding is constant across time for a given receiver, fusing it into every frame-level token enables the attention layers to condition temporal evidence on explicit spatial priors at each timestep; this encourages the model to learn time-varying observation-to-geometry mappings and yields improved temporal consistency and pose accuracy.

\rev{\textbf{Spatio-temporal Transformer.}
The Transformer encoder we used consists of $L=6$ stacked layers. Each layer comprises two sub-layers: a Multi-Head Self-Attention (MSA) layer and a Feed-Forward Network (FFN). Layer Normalization (LN) is applied before the input of each sub-layer, followed by a residual connection. The MSA utilizes $8$ heads with a feature dimension of $512$. The FFN consists of two linear transformations with GeLU activation and a dropout rate of $0.1$, expanding the intermediate dimension to $2048$. }

%\paragraph{Decoding and pose estimation.}This architecture effectively captures global spatio-temporal dependencies across the token sequence.
%The token sequence \(\mathcal{U}^{(0)}\) is processed by a Transformer-based encoder with \(L\) layers of multi-head self-attention:
%\begin{equation}
%\mathcal{U}^{(\ell)} \;=\; \mathcal{TF}^{(\ell)}\big(\mathcal{U}^{(\ell-1)}\big),\qquad \ell=1,\dots,L,
%\end{equation}
%producing contextualized tokens \(\mathbf{u}_{n,t}^{(L)}\). For each frame \(t\) the \(N_r\) tokens \(\{\mathbf{u}_{n,t}^{(L)}\}_{n=1}^{N_r}\) are aggregated to produce a frame-level feature \(\mathbf{z}_t\in\mathbb{R}^D\). A pose head \(h_\phi\) then regresses the predicted joints:
%\begin{equation}
%\hat{\mathbf{y}}_t \;=\; h_\phi(\mathbf{z}_t) \in \mathbb{R}^{J\times 3}.
%\end{equation}

\rev{\textbf{Decoding and pose estimation.}
The input \(\mathcal{U}^{(0)}\) is first processed by the spatiotemporal Transformer encoder, which produces contextualized tokens denoted as \(\mathbf{u}_{n,t}^{(L)}\). Then, to synthesize multi-view information, we reshape this output. Specifically, for each frame \(t\), the \(N_r\) tokens \(\{\mathbf{u}_{n,t}^{(L)}\}_{n=1}^{N_r}\) are combined to form a frame-level representation \(\mathbf{z}_t \in \mathbb{R}^{N_r \times D}\), where $D$ denotes the token dimension. A pose head decoder \(h_\phi\) then regresses the predicted joints:
\begin{equation}
\hat{\mathbf{y}}_t \;=\; h_\phi(\mathbf{z}_t) \in \mathbb{R}^{J\times 3}.
\label{eq:pose_head}
\end{equation} 
The pose head decoder \(h_\phi\) is a lightweight MLP designed to map the final feature \(\mathbf{z}_t\) to skeletal keypoint coordinates. Concretely, it progressively reduces the feature dimensionality through a sequence of fully connected layers, following the architecture \(N_r \times D \rightarrow 1024 \rightarrow 512 \rightarrow J \times 3\), where \(J\) denotes the number of skeletal joints.}

%This yields a unified frame-level feature vector \(\mathbf{z}_t \in \mathbb{R}^{N_r \times D}\), which fuses the geometric context from all sensing nodes. Finally, a lightweight MLP decoder \(h_\phi\) (structured as \(N_r D \to 1024 \to 512 \to J \times 3\)) regresses the predicted joints:
%\begin{equation}
%\hat{\mathbf{y}}_t \;=\; h_\phi(\mathbf{z}_t) \in \mathbb{R}^{J\times 3},
%\label{eq:pose_head}
%\end{equation}
%where \(J=25\) denotes the number of skeletal joints. This design efficiently leverages the global spatio-temporal context to map fused features directly to precise 3D Cartesian coordinates.}

\textbf{Loss function.}
We train the network using a simple mean-squared-error (MSE) loss between the predicted and ground-truth skeleton joint coordinates. Let \(\hat{\mathbf{y}}_{l,j}\in\mathbb{R}^3\) and \(\mathbf{y}_{l,j}\in\mathbb{R}^3\) denote the predicted and ground-truth positions of joint \(j\) at video frame \(l\), respectively, with \(l=1,\dots,L\) and \(j=1,\dots,J\). The training objective is
\begin{equation}
\mathcal{L}_{\mathrm{MSE}} \;=\; \frac{1}{LJ}\sum_{l=1}^L\sum_{j=1}^J \big\| \hat{\mathbf{y}}_{l,j} - \mathbf{y}_{l,j} \big\|_2^2.
\end{equation}

Once the system is trained, deploying it in a new \rev{scene} requires only minimal setup. The user places $\mathbb{B}$ at any convenient location and $\mathbb{B}_1$ at the new transceiver pair to complete the position encoding process. By providing these encodings as conditional parameters to the model, the pre-trained system can be directly applied in the new settings.

%The proposed network is a compact, end-to-end architecture that explicitly integrates calibrated 3D receiver geometry with per-frame CSI evidence. By combining ResNet-derived per-receiver features, spatial encoding and learnable temporal embeddings into tokens, a Transformer encoder can efficiently learn cross-receiver spatial relations and fine-grained temporal dynamics. This design yields a lightweight yet expressive model that is robust to hardware and scene biases, and that effectively leverages spatial priors to improve 3D pose estimation from WiFi signals.

\section{Datasets}
\label{sec:dataset}

%This section describes the visual–WiFi multi-modal dataset we collected and will release. We cover acquisition hardware and nominal parameters, the recording protocol and action taxonomy, calibration and annotation, synchronization and grouping of CSI into frame-level excerpts, canonical splits for cross-domain evaluation, planned release contents, and a short comparison with existing public datasets.
To validate our approach, we design and collect the largest cross-domain WiFi-based 3D pose estimation dataset, %In this section, we provide an overview of the dataset construction process, its content, and a comparison with existing datasets. 
\rev{and it is available at \url{https://github.com/Trymore-lab/PerceptAlign}.}

\subsection{Dataset overview}
%The corpus contains recordings from 21 participants in three indoor environments. As shwon in Figure~\ref{fig:scenes}, these environments are scene\_1 (empty room), scene\_2 (meeting room), and scene\_3 (office). The dataset was collected over a period of six months, with a total size of 370.66 GB and over 82 hours of recorded video. The total number of synchronized frames in the dataset is \(\mathbf{6{,}369{,}186}\). Each capture contains synchronized RGB video and raw WiFi CSI from three receivers, calibration metadata that registers WiFi devices into the world coordinate system, and multi-view visual 3D joint annotations reconstructed via EasyMocap. The hardware we used are described in Table~\ref{tab:device_specs}, and we used 3 WiFi transceiver in this dataset.
The dataset contains recordings from 21 participants across three indoor \rev{scenes}. As shown in Fig.\ref{fig:scenes}, these rev{scenes} include an empty room, a meeting room, and an office. Data collection spanned six months, resulting in \rev{483.22} GB of recordings and over \rev{98} hours of video. The dataset comprises \rev{7,243,590} synchronized frames. Each capture includes synchronized RGB video and raw WiFi CSI from three receivers, calibration metadata that registers WiFi devices into the world coordinate system, and multi-view 3D joint annotations reconstructed with EasyMocap. More details please refer to Appendix~\ref{sec:dataapp}.

\subsection{Action taxonomy}
We record 18 daily-action categories designed to cover a wide range of limb motions and dynamics. The actions are: (1) left arm stretch, (2) right arm stretch, (3) both-arms stretch, (4) left lateral raise, (5) right lateral raise, (6) left forward lunge, (7) right forward lunge, (8) left side lunge, (9) right side lunge, (10) jump, (11) pick-up, (12) clockwise spin, (13) counterclockwise spin, (14) jumping jack, (15) squat, 
(16) left rotation, (17) right rotation, and (18) directional hops (forward/back/left/right). %And the recording protocol are:
%\begin{itemize}
%  \item Each non-rotation action is performed at five predefined spatial points within the activity area. At each point the subject faces one of three orientations (frontal, $+45^{\circ}$, $-45^{\circ}$). For each (point, orientation) the subject repeats the action 3 times; each repetition lasts 5s.
%  \item Rotation actions (clockwise / counterclockwise) and a few high-variation hops may vary slightly in duration depending on subject habit; these rotation actions are recorded without fixed point/orientation constraints.
%  \item Scene\_3 contains three device-placement configurations (Setups A/B/C). Each setup differs in TX–RX relative distances (the exact layouts are illustrated in the dataset release figures); angles (facing orientations) are kept unchanged across setups. Two subjects were recorded per setup.
%\end{itemize}

\subsection{Calibration and ground-truth}
We register WiFi devices into the world coordinate system using a lightweight coordinate unification method described in Section~\ref{sec:Cross-Modal Coordinate Alignment Strategy}. We recorded the unification result for each placement as a calibrated set of transmitter/receiver coordinates stored in the dataset metadata. Visual ground-truth  are estimated using EasyMocap; frames with EasyMocap/OpenPose confidence < 0.8 are filtered and the remaining annotations are spot-checked manually.

%\subsection{Synchronization, grouping and dataset-level preprocessing}
%Raw CSI packet streams are aligned to video frames as follows. For each recording session we truncate all receiver CSI streams to the shortest stream length \(N_c\) to ensure inter-receiver alignment. Let \(N_f\) denote the number of extracted video frames for the clip (30\,fps in this dataset). We group every $G = \lfloor N_c / N_f \rfloor$ consecutive CSI samples into a single frame-level group; any remainder samples are appended to the final group. After grouping each receiver yields per-frame group of nominal shape \(3\times57\times T\) (antennas × subcarriers × samples). These groups constitute the inputs to models. %These groups constitute the dataset-level inputs; all further per-model preprocessing (CSI-ratio, DFS, resizing to image-like tensors) is described in Sec.~\ref{sec:network}

\subsection{Domain splits}
To facilitate reproducible evaluation, we provide standardized splits under different testing scenarios.
\textbf{Per-scene:} 80/20 train/test split by action instances for each scene.
\textbf{Cross-user:} leave one subject out across 21 participants.
\textbf{Cross-scene:} leave one scene out (train on two scenes, test on the third).
\textbf{Cross-setup (scene\_3):} leave one setup-out among Setup A/B/C.
\textbf{Cross-orientation :} leave one orientation-out among the three orientations. 
\textbf{Cross-Location:} leave one location out among the five capture points in each scene. A detailed description of the partitioning scheme also included in the dataset’s README file.
%Explicit fold files and subject/setup IDs are included with the dataset release.
\begin{table}[ht]
\centering
\vspace{-0.0in}
\caption{Dataset comparison,and PiW means Person-in-WiFi.}
\label{tab:dataset_comparison}
\begin{tabular}{@{}l r r r c@{}}
\toprule
Dataset & \#Subjects & \#Actions & \#Frames & Multi-layout \\
\midrule
MM-Fi & 40 & 16 & 320k & No \\
PiW-3D & 7 & 8 & 97k & No \\
Ours & 21 & 18 & \rev{7243.5k} & Yes \\
\bottomrule
\end{tabular}
\vspace{-0.2in}
\end{table}

\subsection{Dataset comparison}
%Table~\ref{tab:dataset_comparison} compares our dataset against representative prior multi-modal and WiFi-based corpora.  Two direct contrasts are worth highlighting.

%Table~\ref{tab:dataset_comparison} compares our dataset with current large-scale WiFi-based human pose estimation datasets. Compared to Person-in-WiFi-3D \cite{yan2024person} (reported statistics: 7 subjects, $\approx$97k frames), our dataset substantially increases subject diversity and overall scale. We collect data from 21 participants, record a markedly larger total frame count, and cover a wider repertoire of everyday actions. This expanded subject and action coverage reduces overfitting to individual motion styles and yields more reliable estimates of cross-user generalization. Although the number of participants is smaller than in MM-Fi, compared to MM-Fi~\cite{yang2023mm} and Person-in-WiFi-3D, our dataset places greater emphasis on the cross-domain challenges that are most relevant in real-world applications. Unlike these datasets, in our collection the WiFi device layout is different in every scenario, enabling rigorous evaluation under simultaneous environment and layout shifts. We also include multiple layout settings within the same environment to facilitate analysis of layout-induced variations. In addition, our dataset contains extensive coverage of diverse user orientations and positions, supporting broader generalization studies.

Table~\ref{tab:dataset_comparison} compares our dataset with existing large-scale WiFi-based human pose estimation datasets. Compared to Person-in-WiFi-3D~\cite{yan2024person}, our dataset significantly expands both subject diversity and overall scale. This broader coverage mitigates overfitting to individual motion styles and provides a stronger basis for evaluating \rev{cross-subject} generalization. Although the number of participants is smaller than in MM-Fi~\cite{yang2023mm}, our dataset places greater emphasis on the cross-domain challenges most relevant to real-world applications. Unlike prior datasets, which include \rev{cross-scene} data collection but use the same WiFi device layout within each \rev{scene}, our dataset assigns a different device layout to every scenario. This design enables rigorous evaluation under simultaneous \rev{scene} and layout shifts, a condition that is far more common in real-world applications. We also include multiple layouts within the same \rev{scene} to analyze layout-induced variations, and capture extensive data across diverse \rev{subject} orientations and positions to support more comprehensive generalization studies.

Beyond pose estimation, our dataset can also support cross-domain activity recognition research, as it encompasses a wide variety of actions. Beyond scale and diversity, a key advantage of our release is the inclusion of detailed geometric information of the sensing system, along with scene photographs capturing the calibration checkerboards. Our work demonstrates that such geometric information plays a crucial role in enabling the development of more generalizable WiFi sensing systems. We therefore believe that this new dataset, with its comprehensive geometric annotations, will greatly benefit the research community.
%our dataset places a stronger emphasis on within-scene and within-deployment variation: for each scene we record multiple capture points, each with three facing directions (frontal, $+45^\circ$, $-45^\circ$) and several distinct TX/RX placement configurations (Setups A/B/C in scene\_3).  In practice this design produces systematic variation in distance-to-link, occlusion patterns, angular incidence, and Fresnel geometry while keeping the broad scene semantics fixed—an experimental regime that isolates deployment-geometry effects from scene-level differences.

%Beyond scale and variety, a critical advantage of our release is per-sequence geometric registration: every recorded sequence includes board-assisted WiFi-to-camera calibration so that transmitter/receiver coordinates are expressed in the same camera/world frame as the visual ground truth.  This explicit device-to-visual alignment (i) converts a major latent degree of freedom into observed metadata for learning, and (ii) enables rigorous cross-setup and cross-scene evaluations that are not possible with datasets lacking calibrated device coordinates.  Collectively these properties—greater subject/action diversity, fine-grained within-scene deployment variation, and per-sequence coordinate alignment—make our dataset particularly well suited for studying cross-deployment robustness and for training geometry-conditioned WiFi-pose models.

\subsection{Limitations and usage notes}
The dataset targets single-person indoor pose estimation and cross-domain evaluation; it does not include multi-person or outdoor scenes. The effective temporal resolution for pose recovery is bounded by commodity CSI packet rates and by the grouping strategy. Users should consult the provided metadata to select samples with desired temporal granularity. Consent and ethics: participants signed informed-consent forms; visual data are provided under a controlled-access policy (details in the release).

\section{Evaluations}
\label{sec:Experiments}

\begin{table}[h]
\centering
\vspace{-0.1in}
\caption{\rev{Acquisition hardware, software and nominal parameters.}}
\label{tab:hardware}
\resizebox{\linewidth}{!}{%
\begin{tabular}{l|l|l}
\toprule
\textbf{Modality} & \textbf{Hardware} & \textbf{Key Parameters} \\
\midrule
RGB Video & Intel RealSense D435 & $1920\times1080$, 30 fps, RGB-D aligned \\
WiFi CSI & Intel 5300 + CSI Tool & 1 TX, 3 RX (3 antennas/RX); 57 subcarriers \\
& & \rev{Bandwidth: 20 MHz; Freq: 5.2 GHz; Rate: 810 Hz} \\
& & \rev{Antenna: Omnidirectional} \\
Calibration & Checkerboard ($11\times8$) & Square size: 30 mm; EasyMocap-compatible \\
Ground-truth & EasyMocap & Camera-centric 3D skeleton reconstruction \\
\bottomrule
\end{tabular}%
}
\vspace{-0.2in}
\end{table}

\paragraph{\rev{Hardware configuration.}}
\rev{Our hardware system consists of a WiFi and a visual sensing platform, with detailed specifications summarized in Table~\ref{tab:hardware}. The WiFi sensing platform includes one transmitter and three receivers, all equipped with Intel 5300 Network Interface Cards and configured to collect CSI using the picoscenes tool. The transmitter operates with a single antenna, while each receiver is equipped with three antennas. For visual data acquisition, we employ Intel RealSense D435 depth cameras to capture RGB video at a resolution of $1920 \times 1080$ and a frame rate of 30 fps.}

%We construct a synchronized multimodal data acquisition system comprising a WiFi sensing platform and a vision-based ground-truth system. The detailed specifications are listed in Table~\ref{tab:hardware}. The WiFi sensing component utilizes an Intel 5300 Network Interface Card (NIC) equipped with the Linux 802.11n CSI Tool, operating with 1 transmitter (TX) and 3 receivers (RX). Each transceiver is equipped with 3 omnidirectional dipole antennas. The system operates on the 5.2 GHz frequency band with a 20 MHz bandwidth, capturing CSI data across 57 subcarriers at a sampling rate of 810 Hz.For visual ground truth, we employ Intel RealSense D435 depth cameras capturing RGB video at $1920 \times 1080$ resolution and 30 fps.}

\begin{table*}[t]
\centering
\caption{\rev{Comprehensive quantitative comparison across different domain shifts. We report MPJPE (mm, $\downarrow$), PCK@20 (\%, $\uparrow$), and PCK@50 (\%, $\uparrow$).}}
\label{tab:cross_domain_results}
\footnotesize 
\setlength{\tabcolsep}{4pt} 
\renewcommand{\arraystretch}{1.1}
\begin{tabular}{l|ccc|ccc|ccc}
\toprule
\multirow{2}{*}{\textbf{Evaluation Setting}} & \multicolumn{3}{c|}{\textbf{Person-in-WiFi 3D}} & \multicolumn{3}{c|}{\textbf{DT-Pose}} & \multicolumn{3}{c}{\textbf{PerceptAlign (Ours)}} \\
\cmidrule(lr){2-4} \cmidrule(lr){5-7} \cmidrule(lr){8-10}
 & \textbf{MPJPE} $\downarrow$ & \textbf{PCK@20} $\uparrow$ & \textbf{PCK@50} $\uparrow$ & \textbf{MPJPE} $\downarrow$ & \textbf{PCK@20} $\uparrow$ & \textbf{PCK@50} $\uparrow$ & \textbf{MPJPE} $\downarrow$ & \textbf{PCK@20} $\uparrow$ & \textbf{PCK@50} $\uparrow$ \\ 
\midrule
\textbf{In-Domain} & 221.0 & 18.5 & 48.2 & 156.5 & 32.4 & 68.5 & \textbf{137.2} & \textbf{55.2} & \textbf{88.7} \\ 
\midrule
\textbf{Cross-Location} & 253.1 & 14.2 & 41.5 & 220.0 & 21.8 & 55.4 & \textbf{144.6} & \textbf{50.1} & \textbf{86.3} \\ 
\textbf{Cross-Orientation} & 254.0 & 13.8 & 40.8 & 255.7 & 17.5 & 48.9 & \textbf{147.7} & \textbf{49.5} & \textbf{85.1} \\ 
\textbf{Cross-Subject} & 266.7 & 12.1 & 39.5 & 260.5 & 16.2 & 46.5 & \textbf{145.3} & \textbf{51.3} & \textbf{84.8} \\ 
\midrule
\textbf{Cross-Layout} & 649.3 & 2.5 & 10.4 & 583.2 & 5.8 & 18.2 & \textbf{170.2} & \textbf{46.8} & \textbf{80.1} \\ 
\textbf{Cross-Scene} & 717.2 & 1.8 & 8.5 & 571.1 & 6.5 & 20.1 & \textbf{181.5} & \textbf{44.2} & \textbf{79.5} \\ 
\bottomrule
\end{tabular}
\vspace{-0.0in}
\end{table*}

\paragraph{Model settings.}
%All experiments use a fixed training recipe and model configuration. The CNN encoder is a pretrained ResNet-34 whose pooled output is projected to a feature dimension \(D=512\); WiFi transceiver coordinates are encoded with a multi-frequency mapping using \(K=8\) frequency bands and projected to the same dimension. The fusion module is a 6 layers Transformer encoder, with 8 attention heads and hidden size 512. Models are trained end-to-end with Adam (initial learning rate \(1\times10^{-4}\)), a cosine-annealing schedule that decays the learning rate to \(1\times10^{-6}\) over the 200 training epochs, batch size 64, and weight decay \(1\times10^{-5}\). The spatial-embedding projection MLP is initialized with a small scale so that spatial priors do not dominate early training. Training uses an MSE loss on 3D joint coordinates and evaluation is reported in MPJPE (mm). For comparisons we use the Person-in-WiFi-3D baseline (author weights) evaluated on the same splits.
\rev{Our model is implemented using the PyTorch framework. All training and inference processes are conducted on one NVIDIA GeForce RTX 4090 GPU.}
All experiments follow a fixed training recipe and model configuration. 
%The CNN encoder is a pretrained ResNet-34, with its pooled output projected to a 512-dimensional feature space. WiFi transceiver coordinates are encoded using a multi-frequency mapping with $K=8$ frequency bands and similarly projected to dimension 512. Decoding is performed by a 6-layer Transformer encoder with 8 attention heads and a hidden size of 512.
Models are trained end-to-end using Adam with an initial learning rate of $1\times10^{-4}$, decayed to $1\times10^{-6}$ via a cosine-annealing schedule over 200 epochs. The batch size is 64, with weight decay set to $1\times10^{-5}$. Training minimizes MSE loss on 3D skeleton coordinates, and evaluation is reported using MPJPE (mm) \rev{ and PCK@20/50(\%)}.

\rev{To comprehensively evaluate PerceptAlign, we compare it with two state-of-the-art methods:}
\begin{itemize}
    \item \textbf{\rev{Person-in-WiFi 3D (CVPR 2024) \cite{yan2024person}}}: 
    \rev{A supervised framework that employs Transformers for global correlation modeling. We adopt it as a strong baseline due to its demonstrated superiority over prior approaches and the availability of its open-source implementation.}
    
    \item \textbf{\rev{DT-Pose (CVPR 2025) \cite{chen2025towards}}}: 
    \rev{A self-supervised approach that alleviates cross-domain discrepancies through temporally consistent contrastive masking. We include this method to highlight the benefits of our explicit geometry conditioning compared to implicit representation learning in the presence of domain shifts.}
\end{itemize}

\rev{All baseline methods are retrained from scratch on our dataset and fully converged, while strictly adhering to the same training and testing splits as PerceptAlign.}
%For comparison, we evaluate the state-of-the-art method Person-in-WiFi-3D~\cite{yan2024person} (using the author-released weights). We choose this baseline because many recent studies have not released their code, whereas Person-in-WiFi-3D provides open-source implementations and reports performance surpassing prior work.

%For comparison, we evaluate the state-of-the-art method Person-in-WiFi-3D~\cite{yan2024person} (using the author-released weights) on the same splits.The spatial-embedding projection MLP is initialized with a small scale to prevent spatial priors from dominating early training.

\paragraph{Evaluation setup}
\label{sec:evaluation setup}
%We evaluate pose estimation performance using the standard Mean Per-Joint Position Error (MPJPE). Let \(\hat{\mathbf{y}}_{l,j}\in\mathbb{R}^3\) and \(\mathbf{y}_{l,j}\in\mathbb{R}^3\) denote the predicted and ground-truth 3D positions of joint \(j\) at frame \(l\), respectively, with \(l=1,\dots,L\) and \(j=1,\dots,J\). The MPJPE (reported in millimeters) is computed as the average Euclidean distance between predictions and ground truth over all joints and frames:
%\begin{equation}
%\text{MPJPE} \;=\; \frac{1}{LJ}\sum_{l=1}^{L}\sum_{j=1}^{J} \big\| %\hat{\mathbf{y}}_{l,j} - \mathbf{y}_{l,j} \big\|_2.
%\end{equation}
\rev{We employ two standard metrics to evaluate performance:
\begin{itemize}
    \item \textbf{Mean Per Joint Position Error(MPJPE(mm)):} The average Euclidean distance between the predicted and ground-truth joint coordinates.
    \item \textbf{Percentage of Correct Keypoints(PCK@$\sigma$(\%):} The percentage of predicted joints whose Euclidean error is below a threshold $\sigma$. We report PCK@20 and PCK@50, representing the accuracy with strict thresholds of 20mm and 50mm, respectively.
\end{itemize}}

\subsection{In-domain Evaluation}
\label{sec:In-domain evaluation}

%\begin{table}[ht]
%  \centering
%  \caption{In-domain Evaluation (MPJPE (mm)).}
%  \label{tab:overall_accuracy_simple}
%  \begin{tabular}{@{}lc@{}}
%    \toprule
%    Method & MPJPE (mm) \\
%    \midrule
%    Person-in-WiFi-3D  & 221.0 \\
%    PerceptAlign  & \textbf{137.2} \\
%    \bottomrule
%  \end{tabular}
%\end{table}

%\begin{table*}[ht]
%  \centering
%  \caption{Evaluation Results (MPJPE (mm)).}
%  \label{tab:cross_domain_results}
%  \begin{tabular}{@{}lcccccc@{}}
%    \toprule
%    Method & \rev{In-domain}& Location & Orientation & Environment & Layout & User \\
%    \midrule
%    Person-in-WiFi 3D & 221.0 & 253.1 & 254 & 717.2 & 649.3 & 266.7 \\
%    PerceptAlign & \textbf{137.2} & \textbf{144.6} & \textbf{147.7} & \textbf{181.5} & \textbf{170.2} & \textbf{145.3} \\
%    \bottomrule
%  \end{tabular}
%\end{table*}

We report in-domain MPJPE and \rev{PCK@20/50} for models trained and tested within the same scene in Table~\ref{tab:cross_domain_results}. For brevity, we present the overall mean MPJPE and \rev{PCK@20/50} across the three scenes. All reported models share the same training and evaluation protocol. Each scene is split 80/20 by action instances, and errors are averaged across all test frames. \rev{Compared with Person-in-WiFi-3D, PerceptAlign markedly improves in-domain performance by reducing MPJPE from 221.0 mm to 137.2 mm and increasing PCK@20/50 from 18.5/48.2 to 55.2/88.7. Relative to DT-Pose, it further lowers MPJPE from 156.5 mm to 137.2 mm, a 12.3\% improvement, while achieving the same substantial gains in PCK@20/50.} This gain demonstrates that incorporating WiFi device geometry as prior conditional knowledge can significantly improve 3D pose estimation accuracy. We attribute this to the fact that, even though device layouts remain unchanged in in-domain settings, our approach provides the model with explicit awareness of the transceiver configuration. This allows the model to interpret CSI perturbations in a well-defined spatial context, thereby enhancing its ability to map CSI variations caused by human activity to accurate 3D poses.

\subsection{Cross-Domain Evaluation}
\label{sec:Cross-Domain Evaluation}

\begin{figure*}[ht]
  \centering
  \includegraphics[width=1\textwidth]{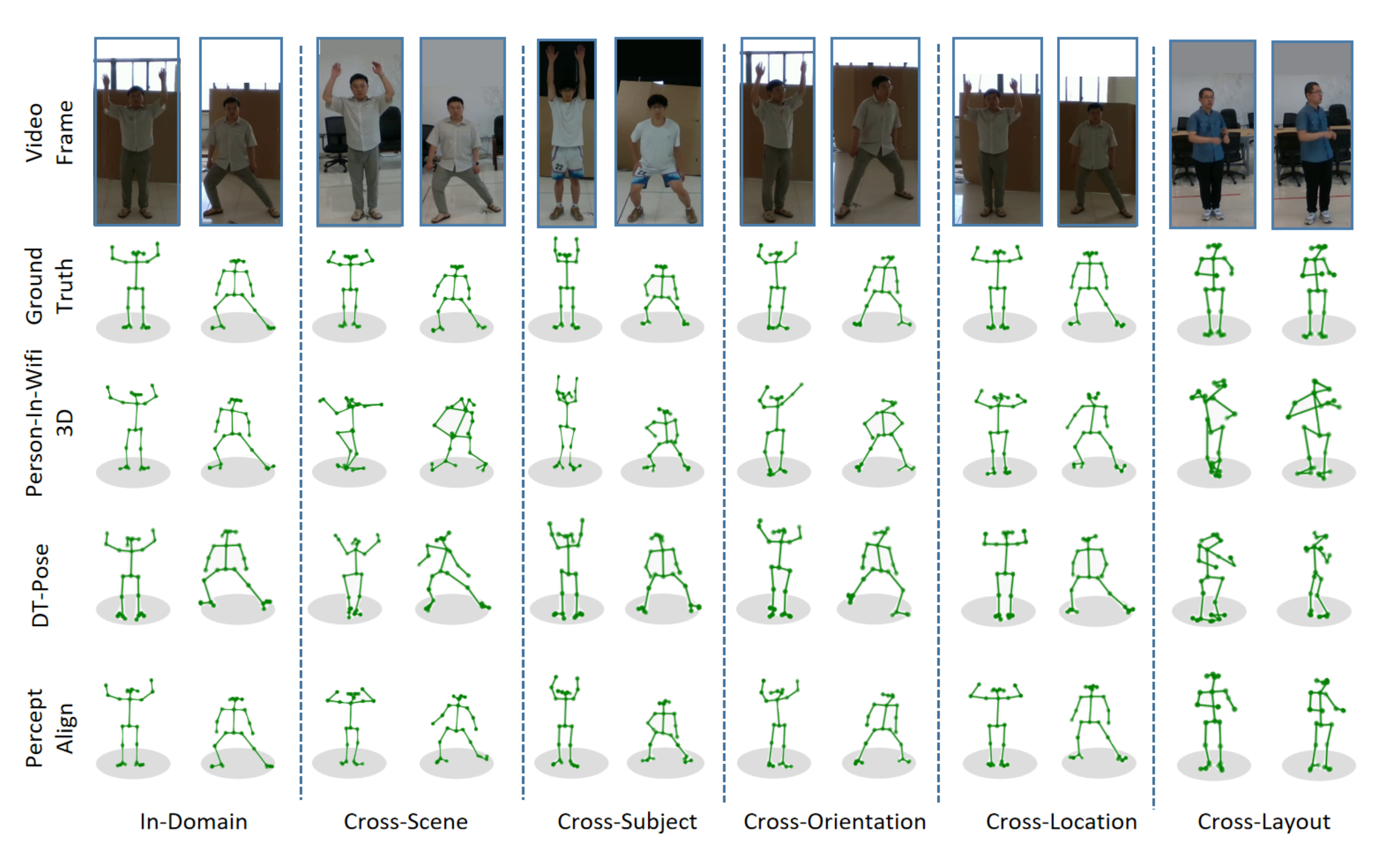}
  \vspace{-0.3in}
  \caption{Illustrative examples of constructed skeletons across different cross-domain settings.}
  \label{fig:In-domain_Cross-domain}
  \vspace{-0.0in}
\end{figure*}

%To evaluate robustness beyond single-deployment accuracy, we conduct cross-domain experiments that probe distinct sources of distribution shift: spatial location, subject orientation, recording environment, subject identity, and device placement. Each protocol follows a leave-one-out design tailored to the respective factor. Table~\ref{tab:cross_domain_results} summarizes results in terms of mean per-joint position error (MPJPE, mm).
To evaluate system performance under realistic deployment variations, we conduct cross-domain experiments targeting different sources of distribution shift: spatial location, subject orientation, \rev{scene}, subject identity, and device layout. Each protocol follows a leave-one-out design, and results are summarized in Table~\ref{tab:cross_domain_results}. And representative samples are shown in Fig.~\ref{fig:In-domain_Cross-domain}.

\textbf{Cross-Location.}  
%This protocol measures how well the model transfers across different subject positions within the same room and device layout. Each capture location is held out in turn, with training on the remaining locations. Person-in-WiFi-3D exhibits marked degradation (221.0\,mm), consistent with sensitivity to link distance and occlusion. By explicitly aligning device coordinates, PerceptAlign reduces the mean error to 145.3\,mm, a relative improvement of 34\%.
This section evaluates how model performance changes when subject positions vary within the same room and device layout. In each round, one position is held out for testing while the remaining positions are used for training. \rev{Compared with in-domain settings, Person-in-WiFi-3D exhibits substantial performance degradation, with MPJPE increasing to 253.1 mm and PCK@20/50 dropping to 14.2/41.5, respectively. DT-Pose also suffers notable degradation, reaching an MPJPE of 220.0 mm with PCK@20/50 of 21.8/55.4. In contrast, although PerceptAlign experiences some performance decline under cross-domain conditions, it significantly mitigates the degradation by incorporating geometric conditioning, reducing the MPJPE to 144.6 mm, which corresponds to a 34.4\% improvement, while raising PCK@20/50 to 50.1/86.3, respectively.
}

%Compared to in-domain settings, Person-in-WiFi-3D shows significant degradation \rev{(MPJPE 253.1 mm, PCK@20/50 14.2/41.5), and DT-pose also shows significant degradation (MPJPE 220.0 mm, PCK@20/50 21.8/55.4)} whereas PerceptAlign also experiences some decline but reduces the MPJPE error to 144.6 mm (43\% improvement) by incorporating geometric conditioning.

\textbf{Cross-Orientation.}  
%To test angular robustness, we adopt a leave-one-orientation-out scheme over frontal, $+45^{\circ}$, and $-45^{\circ}$ views. Baseline performance (254.0\,mm) indicates reliance on orientation specific multipath patterns. PerceptAlign achieves 147.7\,mm, yielding a 42\% reduction in error and demonstrating that geometry-conditioned learning better accommodates directional variation.
For variations in subject orientation, we evaluate performance using a leave-one-orientation-out scheme while keeping other conditions fixed. \rev{Person-in-WiFi-3D attains an MPJPE of 254 mm with PCK@20/50 of 13.8/40.8, while DT-Pose reports an MPJPE of 255.7 mm and PCK@20/50 of 17.5/48.9, indicating that orientation changes substantially hinder generalization. In contrast, PerceptAlign reduces MPJPE to 147.7 mm, corresponding to a 42\%  error reduction, demonstrating its effectiveness under this setting.}%Person-in-WiFi-3D reaches 254 mm, indicating that orientation changes hinder generalization. In contrast, PerceptAlign achieves 147.7 mm, a 42\% reduction in error, demonstrating the effectiveness of our strategy under this setting.

\textbf{\rev{Cross-Subject}.}  
%In leave-one-subject-out testing, the baseline yields 266.7\,mm, revealing poor inter-subject generalization. PerceptAlign reduces error to 145.3\,mm, indicating that disentangling subject motion $h$ from deployment factors $s$ improves transfer across different body shapes and motion styles.
\rev{In the cross-subject evaluation, Person-in-WiFi-3D, DT-Pose, and our method achieve MPJPEs of 266.7 mm, 260.5 mm, and 145.3 mm, respectively. Correspondingly, their PCK@20/50 scores are 12.1/39.5, 16.2/46.5, and 51.3/84.8.} These results highlight the non-negligible impact of subject variation on pose estimation performance, while PerceptAlign maintains robustness through its geometry-conditioned learning strategy. \rev{More details please refer to Appendix~\ref{app:cross-sub}.}

\textbf{Cross-Layout.}  
%Cross-setup evaluation (Table~\ref{tab:cross_domain_results}, ``Setup'') further confirms the benefit of explicit alignment. Person-in-WiFi-3D incurs 649.3\,mm MPJPE, while PerceptAlign lowers the error to 170.2\,mm. Analysis across placements shows that sensitivity to TX–RX spacing and angular coverage is sharply reduced when calibrated device coordinates are exposed to the learner.
The cross-layout setting is a central focus of our work and a key motivation, since layout variations frequently occur in real-world deployments. Results confirm that changes in WiFi device layout severely degrade the performance of existing pipelines: even the state-of-the-art Person-in-WiFi-3D \rev{and DT-Pose} becomes nearly unusable under this setting, \rev{with average MPJPE values of 649.3 mm and 583.2, and PCK@20/50 scores of 2.5/10.4 and 5.8/18.2, respectively.} In contrast, our method achieves \rev{an MPJPE of 170.2 mm and PCK@20/50 scores of 46.8/80.1}, demonstrating strong robustness to layout changes and validating the effectiveness of the proposed geometry-aware approach.

\begin{figure*}[ht]
  \centering
  \includegraphics[width=0.85\linewidth]{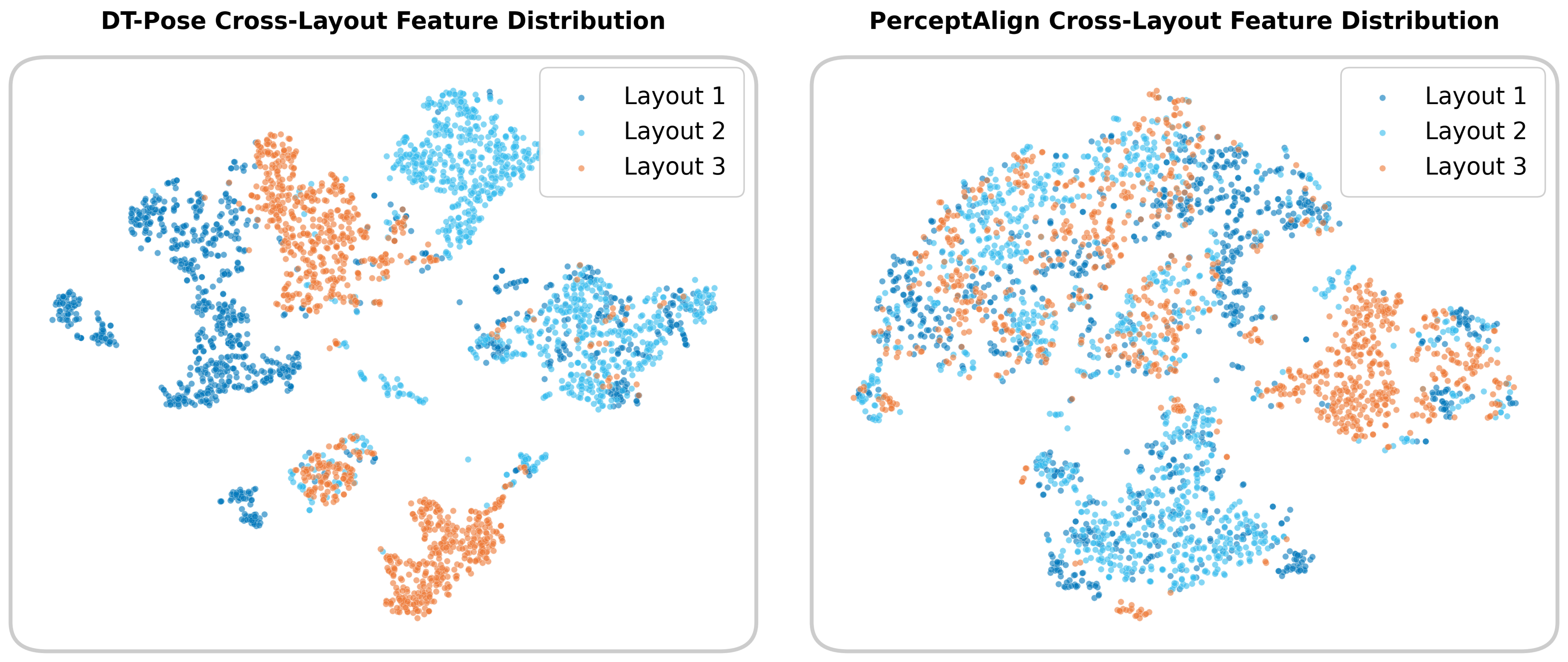} 
  \caption{\rev{t-SNE visualization of feature distributions under the Cross-Layout setting. Blue shades denote training layouts (L1/L2), and orange denotes the unseen test layout (L3).}}
  \label{fig:tsne}
\end{figure*}

\textbf{\rev{Cross-Scene.}}  
%For cross-room generalization we apply leave-one-scene-out across three environments. The baseline suffers severe degradation (717.2\,mm), reflecting strong dependence on environment-specific multipath signatures. PerceptAlign substantially improves robustness, reducing error to 181.5\,mm. We further evaluate leave-one-setup-out within scene\_3, where the baseline yields 649.3\,mm versus 170.2\,mm for PerceptAlign. This highlights that explicit device-to-camera alignment directly addresses deployment-induced failures.
Our \rev{cross-scene} evaluation is more challenging, as it jointly considers changes in both indoor scenes and device layouts—variations that are common in practical deployments. Results show that Person-in-WiFi-3D \rev{and DT-Pose} suffers severe degradation,\rev{with average MPJPE values of 717.2 mm and 571.1 mm, and PCK@20/50 scores of 1.8/8.5 and 6.5/20.1, respectively.}. Although PerceptAlign also experiences some decline, \rev{it maintains an MPJPE of 181.5 mm and PCK@20/50 scores of 44.2/79.5}, demonstrating substantially improved robustness in pose estimation.

%Overall, these results demonstrate that the dominant failure mode of prior work—memorization of deployment-specific couplings—can be alleviated by explicitly conditioning on calibrated device geometry. PerceptAlign consistently reduces cross-domain errors by more than 60\% relative to Person-in-WiFi-3D across all tested protocols.The baseline (Left) clusters distinctively by layout, whereas PerceptAlign (Right) aligns features into a unified space independent of deployment geometry.
Overall, these results validate that existing methods overfit by implicitly memorizing WiFi device layouts, conflating them with useful motion cues. The more than 60\% improvement over the SOTA method across diverse cross-domain settings further demonstrates the effectiveness of our approach: by making device geometry an explicit condition, the model is able to disentangle deployment factors from human motion, leading to significantly stronger generalization. \rev{Moreover, we compare different methods in terms of model size and inference speed to evaluate their deployment feasibility. Person in WiFi 3D~\cite{yan2024person} contains 20.4 million parameters and achieves an inference speed of 54 FPS. The recent DT Pose~\cite{chen2025towards} employs 34.5 million parameters, with an estimated inference speed of 41 FPS. In comparison, our proposed PerceptAlign consists of 29.71 million parameters. Owing to the incorporation of position and temporal encoding, PerceptAlign introduces a moderate computational overhead and attains an inference speed of 37 FPS. This modest reduction in inference speed enables substantially improved generalization performance, particularly under cross-layout and cross-scene settings.}

\subsection{\rev{Evidence of Coordinate Overfitting}}
\rev{To validate our coordinate overfitting hypothesis and demonstrate the effectiveness of PerceptAlign, we visualize the output features of DT-Pose and PerceptAlign using t-SNE~\cite{maaten2008visualizing} under the cross layout setting of Scene 3. All models are trained on Layouts 1 and 2 and evaluated on the unseen Layout 3. The visualized features are taken from the input to the decoder pose head. The results are shown in Figure~\ref{fig:tsne}. As observed, the features produced by DT-Pose (left) form layout specific clusters, indicating that the model overfits to deployment geometry. As a result, samples from the unseen Layout 3, shown in orange, occupy a disjoint region in the feature space. In contrast, the features learned by PerceptAlign (right) exhibit minimal dependence on the layout, substantially alleviating coordinate overfitting. These results suggest that existing state-of-the-art methods implicitly memorize transceiver geometry, which leads to poor generalization, whereas the geometry conditioned learning strategy adopted by PerceptAlign effectively mitigates this issue. }
%\rev{To empirically verify the hypothesis in this paper, Figure~\ref{fig:tsne} visualizes the latent feature distributions using t-SNE~\cite{maaten2008visualizing}. We conduct the experiment under the Cross-Layout protocol (Scene 3), where the model is trained on Layouts 1 and 2 and tested on the unseen Layout 3. For feature extraction, we select the global feature vector immediately preceding the final regression head for the baseline, and the aggregated frame-level feature \(z_t\) (before the pose head \(h_{\phi}\), see Eq.~\ref{eq:pose_head}) for PerceptAlign. The baseline (Left) features cluster distinctly by layout, confirming that the model implicitly encodes deployment geometry, which leaves the unseen Layout 3 (orange) in a disjoint feature space. Conversely, PerceptAlign (Right) aligns all layouts into a unified manifold where clusters correspond to action semantics rather than geometric configurations, demonstrating the effective disentanglement of human motion from deployment artifacts.}

\subsection{Ablation study}
\label{sec:ablation}

\begin{table}[t]
  \centering
  \caption{Ablation results (MPJPE (mm)). ``No Align'' omits geometry-conditioned spatial position embedding; ``No Spatial PE'' injects raw 3D device vectors instead of high-dimensional spatial embeddings.}
  \label{tab:ablation_results}
  \begin{tabular}{@{}lccc@{}}
    \toprule
    Variant & In-domain & Cross-envir & Cross-layout \\
    \midrule
    PerceptAlign     & \textbf{137.2} & \textbf{181.5} & \textbf{170.2} \\
    No Align       & 279.0          & 729.5          & 687.0          \\
    No Spatial PE  & 297.2          & 744.0          & 692.3          \\
    \bottomrule
  \end{tabular}
\vspace{-0.1in}
\end{table}

\begin{figure}[t]
  \centering
  \includegraphics[width=0.85\linewidth]{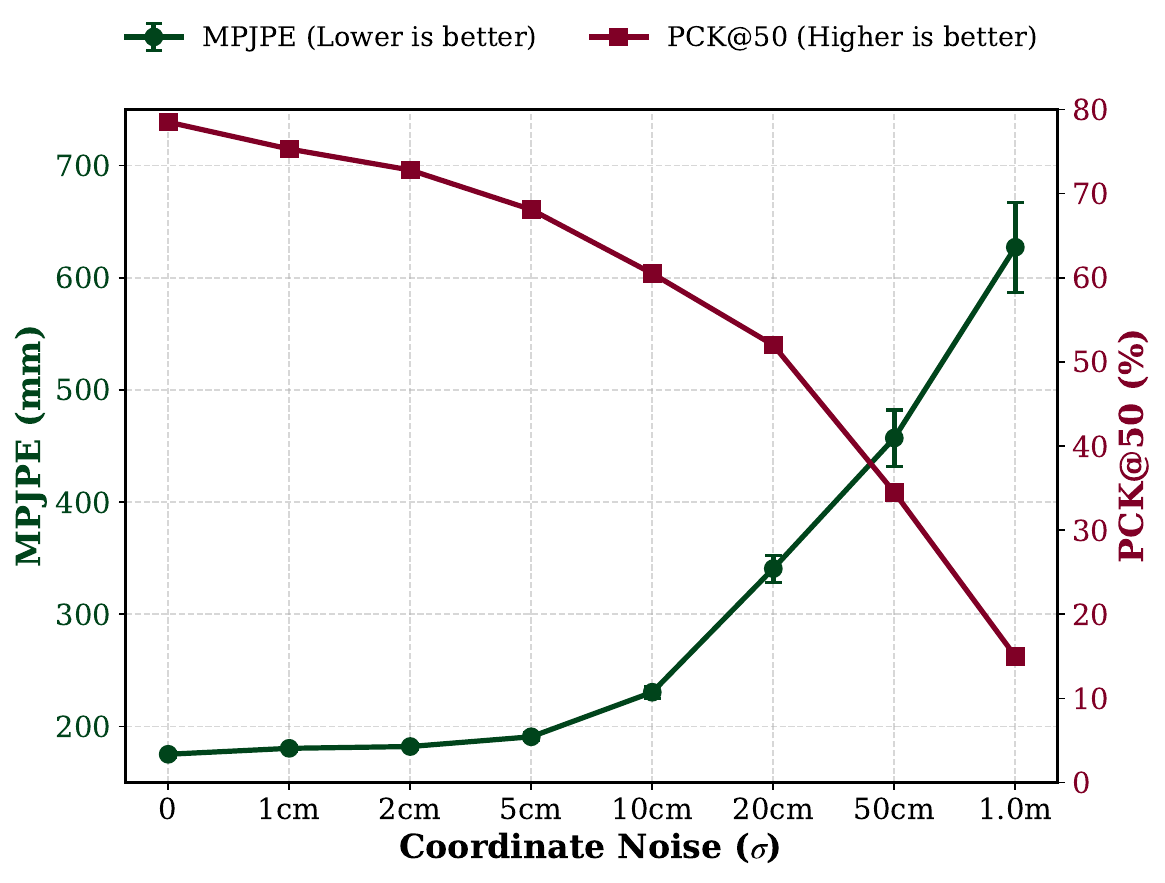} 
  \caption{\rev{Influence of coordinate estimation errors.}}
  \label{fig:sensitivity}
  \vspace{-0.1in}
\end{figure}

Table~\ref{tab:ablation_results} evaluates the contribution of the two core components of PerceptAlign: the geometry-conditioned spatial embedding and the spatial positional encoding (PE) method, which lifts 3D device coordinates into a higher-dimensional representation before fusion.

\textbf{1) Coordinate unification.}
Removing the geometry-conditioned spatial embedding, i.e., training without calibrated device layouts, increases in-domain MPJPE from 137.2 mm to 279.0 mm and cross-domain errors by over 300\% (e.g., \rev{cross-scene} 181.5 → 729.5 mm). This shows that omitting device geometry forces the model to overfit to layout-specific cues: it can fit a single deployment but fails catastrophically under deployment shifts. These results confirm the necessity of incorporating device geometry as explicit conditional knowledge to achieve robust generalization.

\medskip
\textbf{2) High-dimensional spatial encoding.}
Replacing the high-dimensional spatial positional encoding with raw 3D coordinate concatenation leads to severe degradation, with in-domain MPJPE rising from 137.2 mm to 297.2 mm and cross-scene error from 181.5 mm to 744.0 mm. This shows that simply appending coordinates as low-dimensional side information is ineffective: raw vectors lack the nonlinear basis to capture spatial patterns and disrupt feature extraction due to scale mismatch with CSI features. In contrast, lifting coordinates into high-dimensional embeddings provides richer representations that fuse smoothly with CSI, enabling the model to learn geometry-conditioned patterns and maintain robustness under deployment shifts.

%The above experiments clearly demonstrate the effectiveness of the geometry-conditioned spatial embedding and spatial positional encoding schemes. Omitting either component reduces the system to a brittle, deployment-specific pseudo-inverse that collapses under realistic cross-deployment shifts.

%Both ingredients are necessary and complementary. Registering device coordinates \(P\) into the camera frame removes the principal latent degree of freedom that drives coordinate-level overfitting; the high-dimensional spatial PE then converts those coordinates into a form that the fusion network can use effectively to modulate CSI features. Omitting either component returns the system to a brittle, deployment-specific pseudo-inverse that fails under realistic cross-deployment shifts.

\subsection{\rev{Influence of Coordinate Errors}}
\label{sec:sensitivity_analysis}

\rev{Our lightweight coordinate unification procedure may introduce inaccuracies when estimating the positions of WiFi transceivers. To quantify the impact of such errors on system performance, we conduct a sensitivity analysis under the cross layout setting. Specifically, we inject random perturbations with magnitude $\sigma$ into the three dimensional coordinates of the WiFi transceivers, simulating coordinate estimation errors ranging from the centimeter scale ($\sigma = 1$ cm) to the meter scale ($\sigma = 1.0$ m). The results are shown in Figure~\ref{fig:sensitivity}. The results indicate that the proposed model is highly tolerant to centimeter level coordinate errors, while performance degradation becomes pronounced once the error exceeds the decimeter scale. Notably, when $\sigma = 1.0$ m, the MPJPE approaches that of the baselines, suggesting that when geometric priors become unreliable, the model effectively degenerates into a geometry agnostic configuration. In practical deployments, coordinate unification is typically performed using checkerboards that are widely adopted for precise camera calibration. As a result, the induced coordinate estimation errors are usually small and have a negligible impact on overall system performance.}
%\rev{One core premise of PerceptAlign is the explicit conditioning on transceiver geometry. While our lightweight unification system is designed for ease of use, practical deployments may incur measurement errors. To evaluate the robustness of our system against such geometric inaccuracies, we conducted a sensitivity analysis in the most challenging cross-layout setting.Specifically, we injected independent Gaussian noise $\epsilon \sim \mathcal{N}(0, \sigma^2)$ into the 3D coordinates of the WiFi transceivers, simulating calibration errors ranging from millimeter-level ($\sigma=1\text{cm}$) to meter-level ($\sigma=1.0\text{m}$).}

%\rev{The results are presented in Figure~\ref{fig:sensitivity}. The model exhibits high tolerance to coordinate noise up to $\sigma=5\text{cm}$. The MPJPE remains stable, with negligible performance degradation compared to the noise-free baseline. This confirms that PerceptAlign does not require sub-centimeter calibration precision, making it feasible for manual deployment using standard tools (e.g., tape measures) where errors are typically within this safety margin. As the noise magnitude exceeds 10cm, performance degrades monotonically. Notably, at $\sigma=1.0\text{m}$, the MPJPE approaches the performance of the ``No Align'' baseline. This indicates that when geometric priors become completely unreliable, the model effectively reverts to a geometry-agnostic state.}

\subsection{\rev{More Generalization Evaluation}}
\label{sec:novel_structures}

%\rev{To strictly validate the generalization of PerceptAlign across unseen spatial structures, we collected a dedicated test set in two distinct environments: \textbf{Scene 4 (Workstation)} and \textbf{Scene 5 (Corridor)}, as illustrated in Figure~\ref{fig:new_scenes}. These scenes introduce significant structural shifts greatly compared to Scenes 1--3. Furthermore, to demonstrate broader generalization, we used fps=22 for test set collection, compared to the previous fps=30. We adhered to the standard evaluation protocol (3 subjects, 5 locations, 3 orientations) for this testing phase.}

\begin{figure}[t]
    \centering
    \begin{subfigure}{0.48\linewidth}
        \centering
        \includegraphics[width=\linewidth]{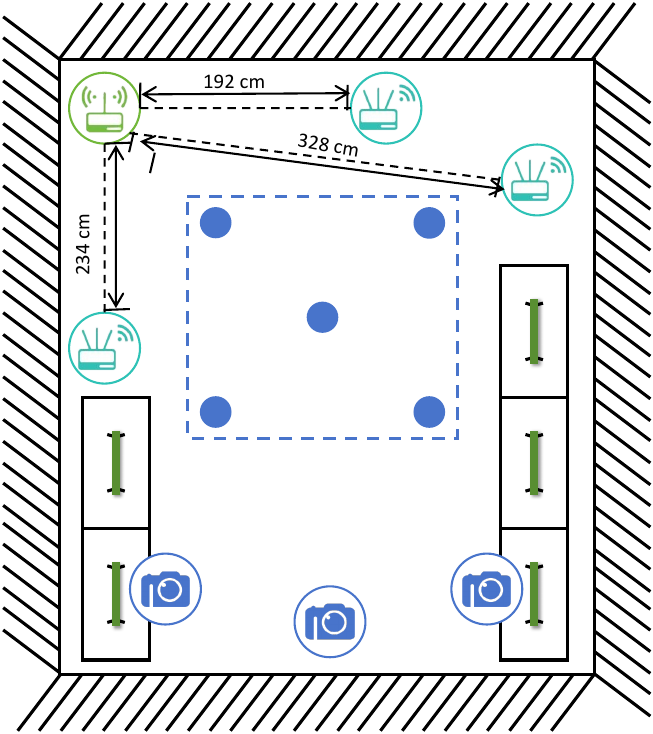}
        \caption{Scene 4: Workstation}
        \label{fig:scene4}
    \end{subfigure}
    \hfill
    \begin{subfigure}{0.48\linewidth}
        \centering
        \includegraphics[width=\linewidth]{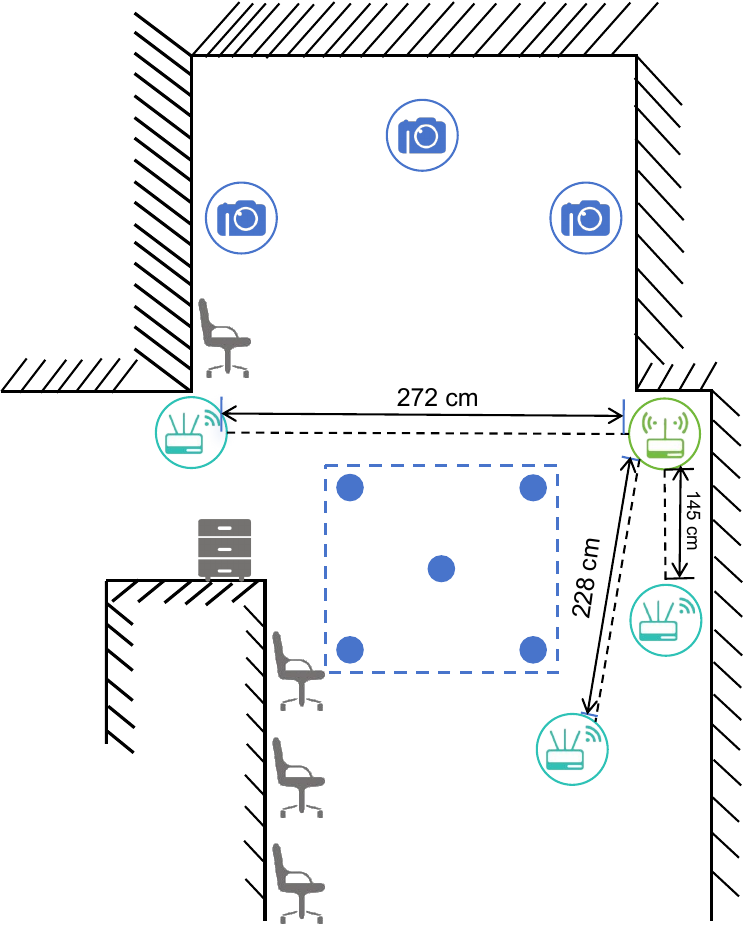} 
        \caption{Scene 5: Corridor}
        \label{fig:scene5}
    \end{subfigure}
    \caption{\rev{Visual illustration of testing scenes.}}
    \label{fig:new_scenes}
      \vspace{-0.1in}
\end{figure}

\rev{To further evaluate the generalization capability of PerceptAlign under previously unseen spatial configurations, we collected an additional test set in two distinct environments, namely \textbf{Scene 4 (Workstation)} and \textbf{Scene 5 (Corridor)}, as shown in Figure~\ref{fig:new_scenes}. Compared with Scenes 1-3, these environments introduce substantially larger structural variations. Moreover, to impose a more stringent generalization setting, this test set was captured at a frame rate of 22 fps rather than the 30 fps used in Scenes 1-3. The motion categories in this new evaluation set are consistent with those in the earlier data collection. For each scene, the dataset includes recordings from 3 subjects, 5 locations, and 3orientations.}

\textbf{\rev{Results and Analysis.}} 
\rev{All models were trained using data from the first three scenes and evaluated on the newly collected test set. Table~\ref{tab:novel_structure_results} summarizes the corresponding results. Existing baselines exhibit severe degradation in these unseen scenes, with MPJPEs spiking above 600mm in Scene 4 and exceeding 700mm in the more geometrically challenging Scene 5. This failure stems from their implicit overfitting to the training layouts. In contrast, PerceptAlign maintains strong robustness. Despite the significant structural and fps differences, our method achieves an MPJPE of 222.4 mm in Scene 4 and 317.1 mm in Scene 5, outperforming the state-of-the-art by over 54.3\%. These results demonstrate that explicitly conditioning on transceiver geometry enables the model to effectively generalize to structurally diverse environments.}%Notably, even in the corridor environment where baseline errors skyrocket, PerceptAlign sustains a high PCK@50 of 56.1\%. 

\begin{table}[h]
\centering
\caption{\rev{Performance comparison on Scene 4 \& 5.}}
\label{tab:novel_structure_results}
\resizebox{\columnwidth}{!}{%
\begin{tabular}{l|ccc|ccc}
\toprule
\multicolumn{1}{c|}{\multirow{2}{*}{\textbf{Method}}} & \multicolumn{3}{c|}{\textbf{Scene 4 (Workstation)}} & \multicolumn{3}{c}{\textbf{Scene 5 (Corridor)}} \\ \cmidrule(lr){2-4} \cmidrule(lr){5-7} 
\multicolumn{1}{c|}{} & \textbf{MPJPE} $\downarrow$ & \textbf{PCK@20} $\uparrow$ & \textbf{PCK@50} $\uparrow$ & \textbf{MPJPE} $\downarrow$ & \textbf{PCK@20} $\uparrow$ & \textbf{PCK@50} $\uparrow$ \\ \midrule
Person-in-WiFi 3D & 770.5 & 1.2 & 6.1 & 837.7 & 0.9 & 4.2 \\ 
DT-Pose & 640.1 & 2.8 & 13.5 & 694.3 & 1.5 & 8.7 \\ 
\textbf{PerceptAlign} & \textbf{222.4} & \textbf{39.7} & \textbf{64.1} & \textbf{317.1} & \textbf{32.2} & \textbf{56.1} \\ \bottomrule
\end{tabular}%
}
\end{table}

\section{Related Work}
\label{sec:related}

RF-based human pose estimation fall into (1) specialized radar or mmWave imaging, which yields high spatial resolution but relies on dedicated sensors~\cite{lemoine2025wip,luo2025sliding,an2025pre,ye2024lpformer,ren2024livehps}; and (2) commodity WiFi CSI-based methods, which trade some physical resolution for ubiquity and low cost. Early RF studies demonstrated robust activity recognition and coarse localization using RSSI/CSI features and classical machine learning\cite{wang2015understanding,yan2019wiact,sen2013avoiding}. As research advanced, deep learning models~\cite{li2024uwb,li2025muceiver} began to extract richer spatio-temporal patterns from CSI, enabling finer-grained tasks such as pose estimation~\cite{geng2022densepose,gu2025csipose,jiang2020towards}. %The progression is: activity $\rightarrow$ localization $\rightarrow$ 2D/3D pose estimation or skeleton reconstruction.

(1) Extending from 2D to 3D: Wang et al. \cite{wang2019person} first used Wi-Fi devices to achieve human pose estimation by combining joint heat maps (JHMs) and body part affinity fields (PAFs) from OpenPose to supervise deep learning models. Later, Yan et al. \cite{yan2024person} expanded to 3D pose estimation via a architecture comprising a Wi-Fi encoder, pose decoder, and fine decoder.
(2) Multi-person estimation: Qu et al.  \cite{qu2025multiformer}, Hsu et al. \cite{hsu2024robust}, and Yan et al. \cite{yan2024person} explored multi-user pose estimation via improved resolution, loss functions, and user separation.
(3) Improving accuracy: Huang et al. \cite{huang2025wipe} leveraged the sparsity of joint heat maps and introduced a 3D streaming signal fusion module. Nguyen et al. \cite{nguyen2025robust} proposed an autoencoder denoiser and an estimator focusing on informative OFDM subcarriers. Deng et al. \cite{deng2025csi} applied CSI spatial decomposition to observe spatial and channel-sensitive views. Zhang et al. \cite{zhang2025vst} introduced Vista-Former with dual-stream spatiotemporal attention. Lee et al. \cite{lee2025wi} combined CNNs and transformers for spatiotemporal feature extraction. Gian et al.\cite{d2024hpe} proposed a multi-branch CNN with selective kernel attention, and extended it in \cite{gian2024wilhpe} with a teacher-student framework to enhance resolution efficiently. 
(4) Cross-domain generalization:Several recent papers propose domain-invariant representations, topology- or physics-informed regularizers, or time–frequency fusion strategies to improve robustness. Chen et al.\cite{chen2025towards} learned domain-invariant features with topology constraints. Zhang et al. \cite{zhang2025wivipose} employed cross-layer optimization and bilinear time-spectral fusion. \rev{More recently, advanced physical modeling and domain adaptation techniques have been proposed. RayLoc~\cite{han2025rayloc} introduces a fully differentiable ray-tracing framework to solve localization as an inverse problem. On the other hand, domain adaptation approaches like AdaPose~\cite{zhou2024adapose} and CrossGR~\cite{li2021crossgr} employ alignment losses or source-free adaptation strategies to mitigate distribution shifts. However, these methods typically operate implicitly on feature statistics and often require data from the target domain for adaptation.}

%While achieving high precision, such ray-tracing methods are computationally intensive and require high-fidelity environmental meshes, making them difficult to scale for real-time, portable pose estimation. 
%However, the aforementioned work all relies on vision to provide a fundamental framework for cross-modal supervision of Wi-Fi pose estimation, leading to widespread overfitting issues between the visual and Wi-Fi perspectives. Furthermore, much of this work still requires systematic cross-domain evaluation and richer device metadata. Existing datasets, such as MM-Fi~\cite{yang2023mm} person-in-Wi-Fi 3D~\cite{yan2024person}, suffer from limitations in scale, user diversity, scene complexity, and multiple location and orientation settings within a scene. Most importantly, the lack of unified Wi-Fi-aware device coordinates with vision leads to inherent generalization issues. Our dataset emphasizes a larger number of captured subjects, a larger total frame size, richer scene position/orientation sampling, multiple device position settings, and explicit device-camera calibration metadata, making it a more robust benchmark for future cross-domain evaluation.
However, prior work relies on vision for cross-modal supervision, which causes overfitting between visual and WiFi perspectives and lacks systematic cross-domain evaluation. Existing datasets such as MM-Fi~\cite{yang2023mm} and Person-in-WiFi-3D~\cite{yan2024person} are limited in scale, subject diversity, scene complexity, and within-scene position/orientation variations, and they omit unified WiFi–camera coordinate calibration, leading to poor generalization. Our dataset addresses these gaps by including more subjects, larger frame counts, richer position and orientation sampling, multiple device layouts, and explicit calibration metadata, providing a stronger benchmark for cross-domain evaluation.

\section{Discussion}
\label{sec:discussion}

\textbf{\rev{Deployment Prerequisites.}} \rev{Unlike current methods, PerceptAlign relies on transceiver coordinates to disentangle motion from layout. However, existing public datasets do not provide such coordinate information, which makes evaluation on current benchmarks, such as MM-Fi~\cite{yang2023mm}, infeasible. Although the deployment environment also requires layout encoding, as detailed in Section~\ref{sec:Cross-Modal Coordinate Alignment Strategy}, our approach only relies on two checkerboards and a camera/smartphone. The user simply captures a photo that contains both checkerboards, after which our program automatically performs the coordinate alignment procedure. The entire process completes in under five minutes and introduces minimal user efforts, thereby maintaining high usability.}

\textbf{\rev{Geometric Constraints.}}\rev{ WiFi sensing captures activity through variations in non-line-of-sight (NLOS) reflections and diffraction effects along the line-of sight-path (LOS). When a transmitter and a receiver are co-located, most of the signal energy propagates directly along the LOS path, while NLOS components carry negligible energy and are only weakly influenced by surrounding motion. Consequently, ambient activities become nearly imperceptible under such conditions. For this reason, existing WiFi sensing studies generally avoid this deployment, and co-located routers and client devices are also uncommon in everyday environments. If multiple receivers are mounted on a single device, our approach treats their spatial encodings as shared and identical.} 
%We further clarify the system's behavior under extreme layout conditions. In scenarios where a transmitter and receiver are co-located, the sensing capability is physically limited by the minimal Fresnel zone coverage, resulting in negligible information gain. Conversely, for a single device equipped with multiple receiving antennas (e.g., a standard 3-antenna array), our model treats them as distinct logical links sharing an identical spatial embedding.

\textbf{Limitations.} Despite these strengths, several limitations remain. First, CSI remains noisy and temporally aliased at commodity packet rates; the method’s temporal resolution is therefore bounded by capture hardware and the grouping strategy used to align CSI with video frames. Second, while positional encoding and Transformer fusion improve generalization, deployment changes that radically alter the set of effective links (for example, extreme TX–RX reconfiguration or highly cluttered scenes) still cause noticeable performance degradation; targeted domain-adaptation or lightweight calibration transfer techniques may be required to fully close this gap (similar challenges have been observed for other WiFi-based systems). \rev{Third, our current design primarily focuses on layout shifts and does not explicitly address individual differences or other domain factors. Attributes such as body shape, user position, and motion style also modulate CSI signals, yet they are physically uncorrelated with transceiver geometry. As a result, a residual performance gap remains in cross domain settings, for example an error of approximately 145 mm in cross subject scenarios.} Finally, the present evaluation focuses on single-person scenarios; extending the approach to robust multi-person 3D human pose estimation raises additional challenges (data association, overlapping multipath), as identified in recent multi-person WiFi work. In practice, these limitations point to several concrete research avenues: (i) increasing CSI temporal resolution or incorporating complementary RF modalities (e.g., ultra-wide-band or mmWave) to better capture fast motions; (ii) developing calibration-light domain adaptation or self-supervised fine-tuning routines so that minimal labeled data in a new deployment suffice to restore performance; and (iii) scaling the dataset and model design to support multi-person scenarios while preserving computational efficiency.

\textbf{Future Work.} In future work, we aim to develop a calibration framework for WiFi sensing systems analogous to the intrinsic and extrinsic calibration of multi-camera systems using checkerboards. For intrinsic calibration, our goal is to eliminate deployment-independent hardware biases so that CSI measurements more closely approximate “ideal physical propagation plus minor noise.” To this end, we plan to experiment with direct transceiver connections via coaxial cables as well as antenna array parameter estimation techniques. For extrinsic calibration, our objective is to rapidly unify different WiFi coordinate systems, similar to how cameras establish a consistent mapping from the world coordinate system to individual camera frames. Possible directions include extending our current checkerboard-based vision-assisted calibration method or leveraging the correlation between CSI and controlled motion trajectories. We believe this line of research is critical for bringing WiFi sensing systems closer to practical deployment. \rev{In addition, we plan to extend PerceptAlign by integrating it with existing generalization techniques, such as adversarial domain generalization, to address individual specific statistical shifts that are independent of the physical layout. This direction is expected to further improve the robustness and generalization capability of the system and to narrow the remaining performance gaps.}

\section{Conclusion}
\label{sec:conclusion}
This work identified coordinate overfitting as the main bottleneck in WiFi-based 3D pose estimation, where models memorize deployment-specific layouts. We introduced PerceptAlign, a geometry-conditioned framework that unifies WiFi and vision into a shared 3D space and encodes transceiver positions as priors, disentangling motion from deployment artifacts and enabling layout-invariant features. We also built the largest ross-domain 3D WiFi-based human pose estimation dataset and showed that PerceptAlign reduces in-domain error by \rev{12.3\%} and cross-domain error by over 60\% compared to state-of-the-art baselines. Future work will explore standardized intrinsic and extrinsic calibration protocols to support scalable deployment.% Geometry-aware conditioning thus provides a principled path toward practical WiFi sensing, with broader implications for gesture recognition, activity analysis, and health monitoring.
\begin{acks}
Thanks...
\end{acks}

\bibliographystyle{ACM-Reference-Format}
\bibliography{sample-base}

\appendix

\begin{table*}[htp]
\centering
\caption{\rev{Detailed per-subject MPJPE (mm) for cross-subject evaluation.}}
\label{tab:detailed_cross_user}
\resizebox{\textwidth}{!}{%
\begin{tabular}{c|l|ccccccccccccc|c}
\toprule
\textbf{Scene} & \textbf{Method} & \textbf{Sub1} & \textbf{Sub2} & \textbf{Sub3} & \textbf{Sub4} & \textbf{Sub5} & \textbf{Sub6} & \textbf{Sub7} & \textbf{Sub8} & \textbf{Sub9} & \textbf{Sub10} & \textbf{Sub11} & \textbf{Sub12} & \textbf{Sub13} & \textbf{Avg.} \\ \midrule
\multirow{3}{*}{\textbf{Scene 1}} 
& PiW 3D & 226.7 & 240.2 & 244.8 & 216.8 & 231.0 & 246.3 & 233.9 & 239.4 & 220.6 & 266.3 & 234.0 & 243.3 & 243.0 & \textbf{237.4} \\
& DT-Pose & 248.8 & 255.9 & 222.5 & 221.7 & 238.1 & 229.5 & 236.7 & 203.0 & 228.4 & 230.3 & 228.6 & 229.5 & 215.0 & \textbf{231.4} \\
& \textbf{Ours} & \textbf{131.8} & \textbf{128.7} & \textbf{132.2} & \textbf{122.1} & \textbf{129.0} & \textbf{124.6} & \textbf{133.4} & \textbf{121.6} & \textbf{135.7} & \textbf{128.6} & \textbf{131.0} & \textbf{134.1} & \textbf{132.1} & \textbf{129.6} \\ \midrule

\multirow{3}{*}{\textbf{Scene 2}} 
& PiW 3D & 275.1 & 277.3 & 289.5 & 271.1 & 294.5 & 275.4 & 266.9 & 254.8 & 267.8 & - & - & - & - & \textbf{274.7} \\
& DT-Pose & 280.4 & 248.7 & 292.2 & 275.9 & 261.6 & 261.7 & 255.5 & 272.6 & 262.0 & - & - & - & - & \textbf{267.8} \\
& \textbf{Ours} & \textbf{138.9} & \textbf{151.7} & \textbf{139.1} & \textbf{148.7} & \textbf{151.5} & \textbf{144.2} & \textbf{144.4} & \textbf{148.2} & \textbf{154.8} & - & - & - & - & \textbf{146.8} \\ \midrule

\multirow{3}{*}{\textbf{Scene 3}} 
& PiW 3D & 334.7 & 307.7 & 309.9 & 299.6 & 321.2 & 335.8 & - & - & - & - & - & - & - & \textbf{318.1} \\
& DT-Pose & 300.6 & 358.5 & 313.3 & 316.8 & 298.0 & 308.3 & - & - & - & - & - & - & - & \textbf{315.9} \\
& \textbf{Ours} & \textbf{177.4} & \textbf{168.8} & \textbf{176.5} & \textbf{182.7} & \textbf{184.3} & \textbf{172.2} & - & - & - & - & - & - & - & \textbf{177.0} \\ \midrule
\multicolumn{15}{r|}{\textbf{Global Weighted Average:}} & \textbf{145.3} \\ \bottomrule
\end{tabular}%
}
\end{table*}

\begin{table*}[htp]
\centering
\caption{\rev{Detailed per-subject PCK@20 / PCK@50 (\%) for cross-subject evaluation.}}
\label{tab:detailed_cross_user_pck}
\resizebox{\textwidth}{!}{%
\begin{tabular}{c|l|ccccccccccccc|c}
\toprule
\textbf{Scene} & \textbf{Method} & \textbf{Sub1} & \textbf{Sub2} & \textbf{Sub3} & \textbf{Sub4} & \textbf{Sub5} & \textbf{Sub6} & \textbf{Sub7} & \textbf{Sub8} & \textbf{Sub9} & \textbf{Sub10} & \textbf{Sub11} & \textbf{Sub12} & \textbf{Sub13} & \textbf{Avg.} \\ \midrule
\multirow{3}{*}{\textbf{Scene 1}} 
& PiW 3D & 15.0/47.0 & 13.8/43.9 & 13.4/42.9 & 16.0/49.6 & 14.6/46.0 & 13.3/42.6 & 14.3/45.3 & 13.8/44.1 & 15.6/48.6 & 11.8/38.8 & 14.3/45.3 & 13.5/43.2 & 13.5/43.3 & \textbf{14.1/44.7} \\
& DT-Pose & 16.8/48.0 & 16.1/46.4 & 19.8/54.8 & 19.9/55.1 & 17.9/50.6 & 18.9/52.8 & 18.1/50.9 & 22.8/61.2 & 19.1/53.1 & 18.8/52.6 & 19.0/53.1 & 18.9/52.8 & 20.9/57.1 & \textbf{19.0/53.0} \\
& \textbf{Ours} & \textbf{54.2/88.1} & \textbf{55.8/89.5} & \textbf{54.0/87.9} & \textbf{57.2/91.0} & \textbf{55.6/89.3} & \textbf{56.8/90.2} & \textbf{53.8/87.5} & \textbf{57.5/91.2} & \textbf{52.9/86.8} & \textbf{55.9/89.6} & \textbf{54.8/88.5} & \textbf{53.5/87.2} & \textbf{54.1/87.8} & \textbf{55.1/88.8} \\ \midrule

\multirow{3}{*}{\textbf{Scene 2}} 
& PiW 3D & 11.2/37.3 & 11.1/36.9 & 10.4/35.1 & 11.5/38.0 & 10.1/34.4 & 11.2/37.2 & 11.7/38.7 & 12.6/40.9 & 11.7/38.5 & - & - & - & - & \textbf{11.3/37.4} \\
& DT-Pose & 14.0/41.5 & 16.8/48.0 & 13.2/39.5 & 14.4/42.4 & 15.6/45.2 & 15.5/45.1 & 16.1/46.5 & 14.6/43.0 & 15.5/45.1 & - & - & - & - & \textbf{15.1/44.0} \\
& \textbf{Ours} & \textbf{53.3/88.1} & \textbf{46.7/79.3} & \textbf{53.2/88.0} & \textbf{48.1/81.2} & \textbf{46.8/79.4} & \textbf{50.4/84.2} & \textbf{50.3/84.1} & \textbf{48.4/81.5} & \textbf{45.3/77.4} & - & - & - & - & \textbf{49.2/82.6} \\ \midrule

\multirow{3}{*}{\textbf{Scene 3}} 
& PiW 3D & 8.4/29.5 & 9.5/32.6 & 9.4/32.3 & 9.9/33.7 & 8.9/31.0 & 8.3/29.4 & - & - & - & - & - & - & - & \textbf{9.1/31.4} \\
& DT-Pose & 12.6/38.2 & 9.7/30.9 & 11.9/36.4 & 11.7/35.9 & 12.8/38.6 & 12.2/37.1 & - & - & - & - & - & - & - & \textbf{11.8/36.2} \\
& \textbf{Ours} & \textbf{36.9/65.7} & \textbf{39.8/69.7} & \textbf{37.2/66.1} & \textbf{35.4/63.4} & \textbf{34.9/62.8} & \textbf{38.6/68.1} & - & - & - & - & - & - & - & \textbf{37.1/66.0} \\ \midrule
\multicolumn{15}{r|}{\textbf{Global Weighted Average:}} & \textbf{51.3/84.8} \\ \bottomrule
\end{tabular}%
}
\vspace{-0.1in}
\end{table*}

\section{Appendix A: More details of our dataset}
\label{sec:dataapp}
The hardware specifications are listed in Table\ref{tab:hardware}, and the recording protocol are as follows:
\begin{itemize}
  \item Each non-rotation action is performed at five predefined spatial points within the activity area. At each point the subject faces one of three orientations (frontal, $+45^{\circ}$, $-45^{\circ}$). For each (point, orientation) the subject repeats the action 3 times; each repetition lasts 5s.
  \item Rotation actions (clockwise / counterclockwise) and a few high-variation hops may vary slightly in duration depending on subject habit; these rotation actions are recorded without fixed point/orientation constraints.
  \item Scene\_3 contains three device-placement configurations (Setups A/B/C). Each setup differs in TX–RX relative distances (the exact layouts are illustrated in the dataset release figures); angles (facing orientations) are kept unchanged across setups. Two subjects were recorded per setup.
\end{itemize}
%Raw CSI traces (per receiver, per session) and per-session metadata (timestamps, packet indices). Calibration metadata: recorded checkerboard poses, computed transform matrix, and unified TX/RX coordinates .

The public release included:
\begin{itemize}
  \item Raw CSI traces and per-session metadata.
  \item Synchronized RGB video clips and timestamps.
  \item Calibration metadata: recorded checkerboard poses, computed transform matrix, and unified TX/RX coordinates.
  \item Ground-truth 3D human pose annotations from EasyMocap.
  \item Standardized domain split definitions and data-loading / synchronization scripts.
  \item Illustrative figures showing device-placement schematics for Setups A/B/C in scene\_3.
  \item Code of PerceptAlign and a list of filtered frames.
\end{itemize}

\section{Appendix B: Detailed Cross-Subject Evaluation}

\label{app:cross-sub}

\rev{This section provides additional details on the cross subject evaluation presented in Section 6.2. Table~\ref{tab:detailed_cross_user} and Table~\ref{tab:detailed_cross_user_pck} reports the MPJPE and PCK@20/50 for each subject across all scenes. As shown, all methods achieve their best performance in Scene~1, where the training data exhibits the greatest diversity. Performance degrades in Scene~2 and reaches its lowest level in Scene~3. The baseline methods suffer a pronounced drop in Scene~3, with errors exceeding 300~mm in several cases, indicating that models struggle to generalize when training data is limited. Although PerceptAlign also exhibits increased error in Scene~3, it consistently maintains substantially lower errors than the baselines. This result demonstrates that geometry aware spatial embeddings that leverage layout priors enable the model to learn more effective and transferable features.}
%\rev{This section details the cross-subject evaluation results discussed in Section 6.2. Table~\ref{tab:detailed_cross_user} presents the per-subject MPJPE (mm) across all scenes, while Table~\ref{tab:detailed_cross_user_pck} provides the corresponding PCK@20 and PCK@50 metrics.}

%\rev{As evidenced in Table~\ref{tab:detailed_cross_user}, all methods exhibit optimal performance in Scene~1, which contains the most diverse training data. Performance degrades in Scene~2 and reaches its nadir in Scene~3. Notably, baseline methods suffer severe deterioration in Scene~3, with errors frequently exceeding 300~mm. This trend indicates a limited capacity to generalize under constrained training data regimes.}

%\rev{The PCK results in Table~\ref{tab:detailed_cross_user_pck} corroborate these findings, highlighting the impact of domain shifts on estimation reliability. In the challenging Scene~3, baseline methods show a precipitous drop in precision, with PCK@20 scores often falling below 10, signaling a failure to recover accurate joint locations. Conversely, PerceptAlign maintains robust usability, consistently achieving PCK@50 scores above 60 even in the most demanding scenarios. These results confirm that our geometry-aware spatial embeddings effectively disentangle motion from deployment artifacts, preserving structural validity where baselines fail.}

\end{document}